\documentclass{birkjour}

\usepackage{mathptmx}       % selects Times Roman as basic font
\usepackage{type1cm}        % activate if the above 3 fonts are
\usepackage{bm}

\usepackage{amsmath}
\usepackage{amssymb}
\usepackage{graphicx}        % standard LaTeX graphics tool
\usepackage{url}
\usepackage[noadjust]{cite}
 \theoremstyle{definition}
 
 \theoremstyle{remark}

 \numberwithin{equation}{section}

\newcommand{\midline}{\hline}
\newcommand{\bottomline}{\hline}
\newcommand{\vect}[1]{\mathbf{#1}}
\newcommand{\gvec}[1]{\mathbf{#1}}
\newcommand{\gveca}{{\mathbf{a}}}
\newcommand{\gvecb}{{\mathbf{b}}}
\newcommand{\gvecc}{{\mathbf{c}}}
\newcommand{\be}{\begin{equation}}
\newcommand{\ee}{\end{equation}}

\usepackage{calligra}
\DeclareMathAlphabet{\mathcalligra}{T1}{calligra}{m}{n}
\DeclareFontShape{T1}{calligra}{m}{n}{<->s*[1.0]callig15}{}
\newcommand{\scriptp}{{\Large \calligra{p}}$\,$}
\begin{document}

%-------------------------------------------------------------------------
% editorial commands: to be inserted by the editorial office
%
%\firstpage{1} \volume{228} \Copyrightyear{2004} \DOI{003-0001}
%\seriesextraline{This is the Concrete Title of this Book\br H.E. R and S.T.C. W, Eds.}
%\firstpage{1}

\title[Subperiodic Groups in Clifford's Geometric Algebra]
 {Representation of Crystallographic Subperiodic Groups in Clifford's Geometric Algebra}

%----------Author 1
\author[E. Hitzer]{Eckhard Hitzer}
\address{%
College of Liberal Arts, Department of Material Science,\\ 
International Christian University,\\
181-8585 Tokyo, Japan}
\email{hitzer@icu.ac.jp}

%\thanks{This work was completed with the support of our \TeX-pert.}
%----------Author 2
\author[D. Ichikawa]{Daisuke Ichikawa}
\address{%
Department of Applied Physics,\\ 
University of Fukui,\\ 
910-8507 Fukui, Japan}
\email{seiunnedved3032@yahoo.co.jp}

% 20H15   Other geometric groups, including crystallographic groups [See also 51-XX, especially 51F15, and 82D25]
% 15A66   Clifford algebras, spinors
% 74N05   Crystals
% 20F55   Reflection and Coxeter groups [See also 22E40, 51F15]
% 74E15   Crystalline structure
% 20F05   Generators, relations, and presentations


\date{November 15, 2012}
\dedicatory{Soli Deo Gloria.}
%%% ----------------------------------------------------------------------

\begin{abstract}
This paper explains how, following the representation of 3D 
crystallographic space groups in Clifford's geometric algebra, it is further possible to similarly 
represent the 162 so called subperiodic groups of crystallography in 
Clifford's geometric algebra. A new compact geometric algebra 
group representation symbol is constructed, which allows to read off the complete set 
of geometric algebra generators. For clarity moreover  
the chosen generators are stated explicitly. The group symbols are based on 
the representation of point groups in geometric algebra by versors 
(Clifford monomials, Lipschitz elements).
\end{abstract}

%%% ----------------------------------------------------------------------
\maketitle
%%% ----------------------------------------------------------------------
%\tableofcontents
% ------------------------------------------------------------------------

\section{Introduction}

Crystals are fundamentally periodic geometric arrangements of atoms, ions or molecules. 
The directed distance between two such elements is a Euclidean vector
in $\mathbb{R}^3$. Intuitively all symmetry properties of crystals
depend on these vectors. Indeed, the geometric product of 
vectors \cite{DL:GAP} combined
with the conformal model of 3D Euclidean space 
\cite{SL:Diss,LL:ConfTrCA,HLR:NewAlgTools,EH:ConfM,HL:invalg}
%(Lie, 1872; Lounesto \& Latvamaa, 1980; Hestenes \textit{et al.}, 2001; Hitzer, 2004; Li, 2008) 
yields
an algebra fully expressing crystal point and space 
groups \cite{JG:tesis,DH:PGSG,HP:crystGA,HP:sym3vec}.
%(Gutierrez, 1996; Hestenes, 2002; Hitzer \& Perwass, 2004,2005). 
Two successive reflections at (non-) parallel planes express (rotations) 
translations, etc. \cite{HC:DGGR,CM:GRDG}.
%(Coxeter, 1934; Coxeter \& Moser, 1980).
This leads to a 1:1 correspondence of geometric objects and 
symmetry operators \cite{HTBY:Carrier}
%(Hitzer \textit{et al.}, 2009) 
with vectors and their products, ideal for 
creating a suite of interactive visualizations using CLUCalc \cite{CP:CLUCalc}
%(Perwass, 2002) 
and OpenGL \cite{HP:sym3vec,HP:TheSGV,HP:SGV2010,HPI:SGV+SubpG,EH:VisPlaneG}.
%(Perwass \& Hitzer, 2005; Hitzer \& Perwass 2006). 
%%% Information from reviewer1: 
Independently it has been shown that the coincidence site lattice transformation groups, important for modeling crystalline interfaces and grain boundaries, can be completely determined by coincidence symmetry reflections through hyperplanes orthogonal to lattice vectors \cite{RAV:CACoinPL,YMZ:StrCSL,YMZ:IndCIHypLat,RAAG:CLHypPlane}. 

For material science the subperiodic space groups \cite{KL:ITE}
%(Kopsky \&  Litvin, 2002) 
in 2D and 3D with only one or two degrees of freedom for translations are also of great interest. This article introduces a new algebraic representation for all subperiodic groups of crystallography, including, for the first time, a complete highly compact multiplicative presentation of the generators for each group. By presentation we mean an explicit representation of group elements. We also introduce a compact new system of subperiodic group symbols that enables one to write down the generators for each group directly from the group symbol. Earlier initial work, reported in \cite{HPI:SGV+SubpG}, is thus completed. 

We begin in Section \ref{sc:PGSG} by explaining the representation of point and space groups in conformal geometric algebra. Then in Section \ref{sc:subp} we show how to construct a new compact geometric algebra 
group representation symbol for subperiodic space groups (frieze groups,
rod groups and layer groups), which allows to read off the complete set 
of geometric algebra generators. For clarity we moreover state 
explicitly what generators are chosen.

\section{Point groups and space groups in Clifford's geometric algebra \label{sc:PGSG}}

\subsection{Cartan-Dieudonn\'{e} and Clifford's geometric algebra \label{sc:Cartan}}

%%%%%%%%%%%%%%%%%%%%%%%%%%%%%%%%%%%%%%%%%% NEW %%%%%%%%%%%%%%%%%%%%

\begin{figure}%[t]
\begin{center}
  \resizebox{0.27\textwidth}{!}{\includegraphics{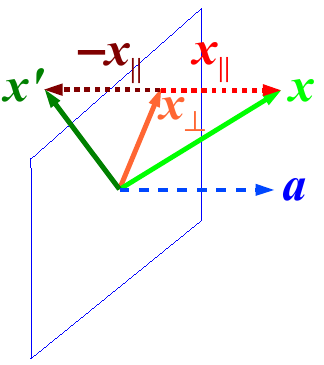}}
  \hspace{2mm}
  \resizebox{0.27\textwidth}{!}{\includegraphics{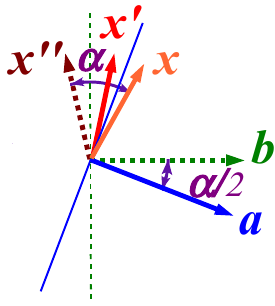}}
  \hspace{2mm}
  \scalebox{0.24}{\includegraphics[width=\textwidth]{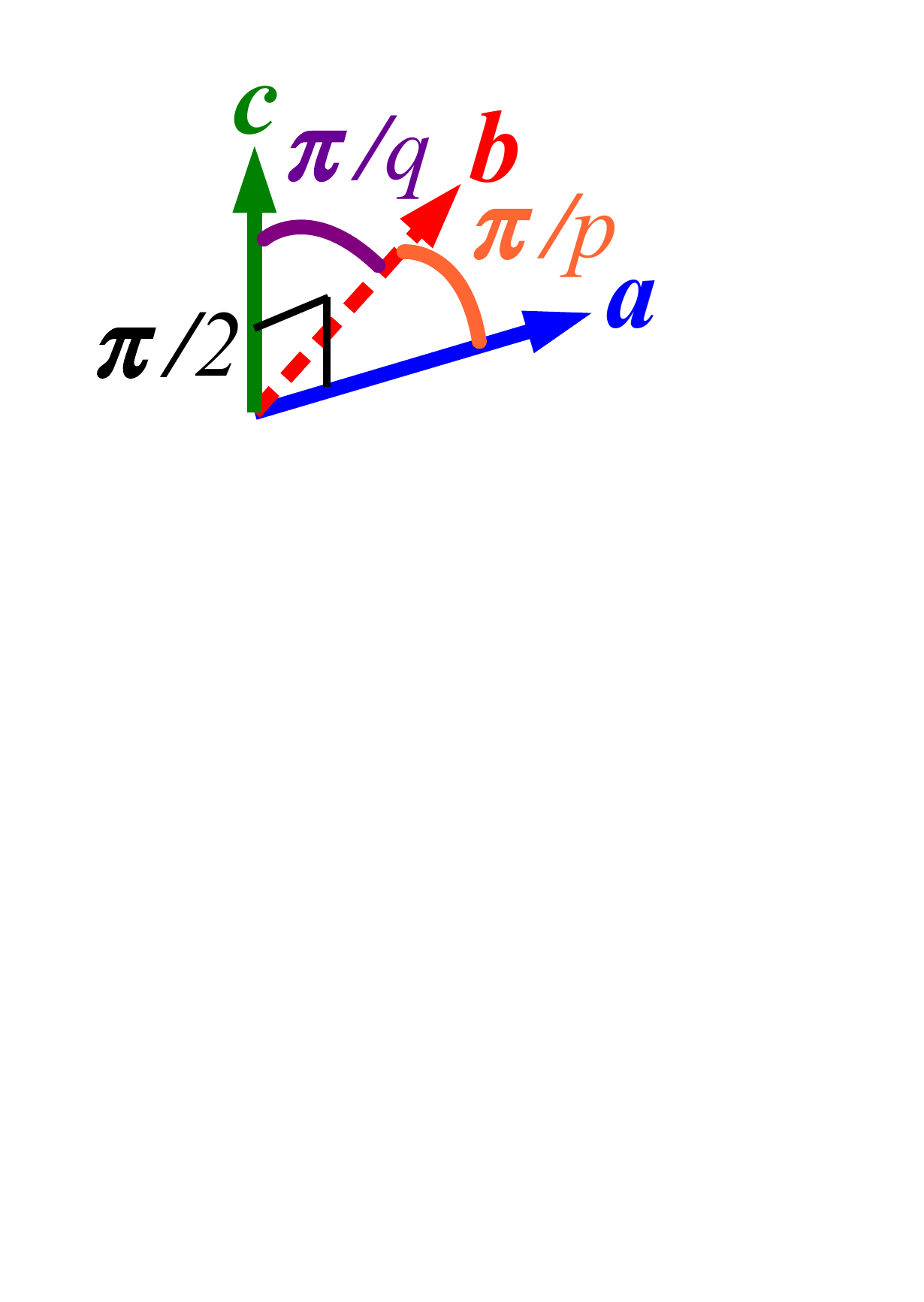}}
  \caption{Left: Reflection at a hyperplane normal to $\gvec{a}$. 
           Center: Rotation generated by two successive reflections 
           at hyperplanes normal
           to $\gvec{a},\gvec{b}$ by twice the 
           angle $\angle(\gvec{a},\gvec{b})$.
           Right: Angular relations of pairs of geometric cell vectors of 
           $\gvec{a},\gvec{b},\gvec{c}$: 
           $\angle(\gvec{a},\gvec{b})=\pi/p$,
           $\angle(\gvec{b},\gvec{c})=\pi/q$,
           $\angle(\gvec{a},\gvec{c})=\pi/2$, $p,q\in \{1,2,3,4,6\}$.
  \label{fg:refrot}}
\end{center}
\end{figure}

Clifford's associative geometric product~\cite{WC:AppGExtAlg,DL:GAP} of two vectors simply adds the (symmetric) inner product to the (anti-symmetric) outer product of Grassmann
\begin{equation}
  \gvec{a}\gvec{b} = \gvec{a}\cdot\gvec{b} + \gvec{a}\wedge\gvec{b}\,.
  \label{eq:gp}
\end{equation}
The mathematical meaning of the left and right side of \eqref{eq:gp} is clear from applying the geometric product to the $n$ orthonormal basis vectors $\{\gvec{e}_1, \ldots , \gvec{e}_n\}$ of the underlying vector space $\mathbb{R}^{p,q}, n=p+q$. We thus have
\begin{align}
  \gvec{e}_k\gvec{e}_k &= 
  \gvec{e}_k\cdot\gvec{e}_k=+1, 
  \quad \gvec{e}_k\wedge\gvec{e}_k=0,\quad 1\leq k \leq p, \\
  \gvec{e}_k\gvec{e}_k &=  
  \gvec{e}_k\cdot\gvec{e}_k=-1,
  \quad \gvec{e}_k\wedge\gvec{e}_k=0, 
  \quad p+1\leq k \leq n, \\
  \gvec{e}_k\gvec{e}_l\, &= -\gvec{e}_l\gvec{e}_k=\gvec{e}_k\wedge\gvec{e}_l,
  \quad \gvec{e}_k\cdot\gvec{e}_l=0,
  \quad l\neq k, \,\, 1\leq k,l \leq n.
\end{align}
Under this product parallel vectors commute and perpendicular vectors anti-commute
\begin{equation}
  \label{eq:cmanticm}
  \gvec{a}\gvec{x}_{\parallel} = \gvec{x}_{\parallel}\gvec{a}\,,
  \quad \quad
  \gvec{a}\gvec{x}_{\perp} = - \gvec{x}_{\perp}\gvec{a}\,.
\end{equation}
This allows to write the \textit{reflection} of a vector $\gvec{x}$ \textit{at a hyperplane} through the origin with normal $\gvec{a}$ as (see left side of Fig. \ref{fg:refrot})
\begin{equation}
  \label{eq:ref}
  \gvec{x}^{\,\,\prime}= -\,\gvec{a}^{\,\, -1} \gvec{x}\,\gvec{a}\, ,
  \quad \quad
   \gvec{a}^{\,\, -1} = \frac{\gvec{a}}{\gvec{a}^{\,\, 2}} 
   \quad \quad
  \gvec{a}^{\,\, -1}\gvec{a}
  = \gvec{a}\gvec{a}^{\,\, -1}
  = 1 \, .
\end{equation}
We can prove \eqref{eq:ref} by beginning with the expression of the reflected vector (see left side of Fig. \ref{fg:refrot} for the meaning of $\gvec{x}_{\parallel}$ and $\gvec{x}_{\perp}$ relative to $\gvec{a}$):
\begin{align}
  \gvec{x}^{\,\,\prime}
  &= -\gvec{x}_{\parallel} + \gvec{x}_{\perp}
  = - \gvec{a}^{\,\, -1}\gvec{a} \gvec{x}_{\parallel} + \gvec{a}^{\,\, -1}\gvec{a}\gvec{x}_{\perp}
  = -\gvec{a}^{\,\, -1}\gvec{x}_{\parallel}\gvec{a}  - \gvec{a}^{\,\, -1}\gvec{x}_{\perp}\gvec{a}
  \nonumber \\
  &= -\gvec{a}^{\,\, -1}(\gvec{x}_{\parallel} + \gvec{x}_{\perp})\gvec{a}
  = -\gvec{a}^{\,\, -1}\gvec{x}\gvec{a},
\end{align}
where we have used \eqref{eq:cmanticm} for the third equality. 

The composition of two reflections at hyperplanes, whose normal vectors $\gvec{a}, \gvec{b}$ subtend the angle $\alpha/2$, yields a rotation around the intersection of the two hyperplanes (see center of Fig. \ref{fg:refrot}) by $\alpha$
\begin{equation}
  \label{eq:rot}
  \gvec{x}^{\,\,\prime\prime}
  = (\gvec{a}\gvec{b})^{-1} \gvec{x}\,\gvec{a}\gvec{b} \, ,
  \quad \quad
  (\gvec{a}\gvec{b})^{-1} = \,\gvec{b}^{\,\, -1}\,\gvec{a}^{\,\, -1} \, .
\end{equation}
Continuing with a third reflection at a hyperplane with normal
$\gvec{c}$ according to the Cartan--Dieudonn\'{e} theorem\footnote{The decomposition of space group transformations in the conformal model $Cl(p+1,q+1)$ according to the Cartan--Dieudonn\'{e} theorem is always possible, though it is clearly not unique, since, e.g., any two Euclidean vectors in the same plane, which enclose the angle $\alpha/2$ will generate the rotation in that plane by the angle $\alpha$. Different choices of lattice vectors allow therefore still to generate the same space group. For example, in the hexagon of Fig. \ref{fg:2Dpg}, vector $\gvec{a}$ could obviously be replaced by any of the other five side vectors and $\gvec{b}$ then by a vector to a respective neighboring vertex.} \cite{EC:Spineurs,JD:GrClass,RAAV:AlgCartDeud,FU:GenOrthMatr} yields rotary reflections (equivalent to rotary inversions with angle $\alpha - \pi$) 
\begin{equation}
 \gvec{x}^{\,\,\prime}
  = -\,(\gvec{a}\gvec{b}\gvec{c})^{-1} \gvec{x}\,\gvec{a}\gvec{b}\gvec{c} \, ,
\end{equation} 
and inversions
\begin{equation} 
 \gvec{x}^{\,\,\prime\prime}
  = -\,i^{-1} \gvec{x} \,i \, ,
  \quad \quad
  i \doteq \gvec{a}\wedge \gvec{b}\wedge \gvec{c},
\end{equation}
where $\doteq$ means equality up to non-zero scalar factors 
(which cancel out in \eqref{eq:symtrafo}). 
In general the geometric product $S=\gvec{a}\gvec{b}\gvec{c}\ldots$ of $k$, normal vectors 
%(the versor $S$) 
corresponds to the composition of reflections to all symmetry transformations~\cite{DH:PGSG,HH:SGinGA} of two-dimensional (2D) and 3D point groups\footnote{Note, that a (geometric) crystal class
contains crystals that share the same type of point group. Therefore, the same symbol is used for both point group and crystal class, but there is a fundamental ontological difference between the two concepts.}, which describe the symmetry of crystal cells, 
\begin{equation}
  \label{eq:symtrafo}
 \gvec{x}^{\,\,\prime}
  = (-1)^k S^{\,-1} \,\gvec{x} \,S
  = \widehat{S}^{\,-1} \,\gvec{x} \,S
  = S^{\,-1} \,\gvec{x} \,\widehat{S} ,
\end{equation}
where $\widehat{S} = (-1)^k S$ is the \textit{grade involution}
or \textit{main involution} in Clifford's geometric algebra (GA). 
We call the product of invertible vectors $S$ in \eqref{eq:symtrafo}
\textit{versor} \cite{HH:SGinGA,DFM:GACS,HL:invalg,HTBY:Carrier},
%(Hestenes \& Holt, 2007; Dorst et al., 2007; Li, 2008; Hitzer \textit{et al.}, 2009), 
but the names \textit{Clifford monomial} of invertible vectors,
\textit{Clifford group element}, or \textit{Lipschitz group element}
are equally in use \cite{PL:CAandSpin,HL:invalg}.
%(Lounesto 2001; Li, 2008).

\subsection{Two dimensional point groups}

\begin{figure}
\vspace*{3mm}
%\begin{center}
\resizebox{0.9\textwidth}{!}{\includegraphics{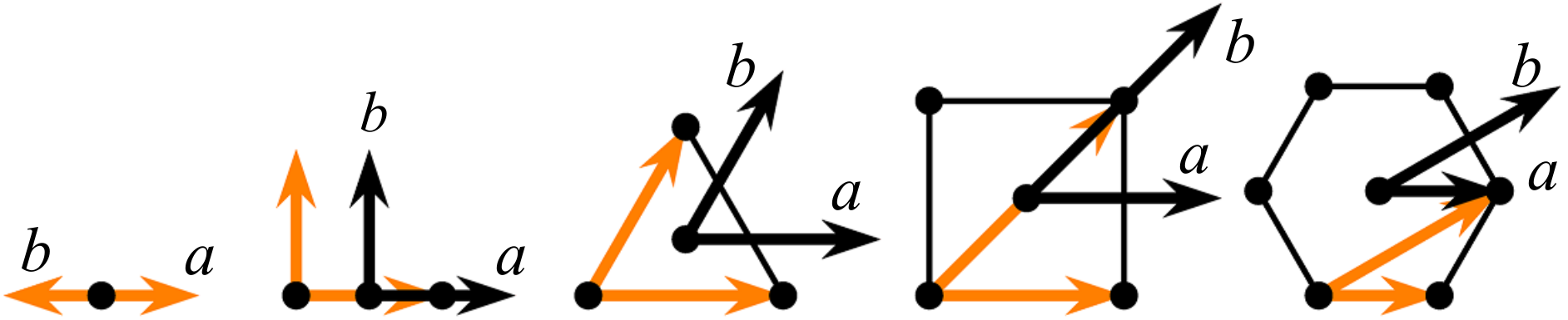}}
%\resizebox{0.9\textwidth}{!}{\includegraphics*[0pt,30pt][542pt,141pt]{twodab2}}
%\resizebox{0.8\textwidth}{!}
%{\includegraphics{twodab.eps}}
\caption{Regular polygons ($p=1,2,3,4,6$) and point 
group generating vectors $\gveca,\gvecb$ subtending angles $\pi/p$ attached to figure center.\label{fg:2Dpg}}
%\end{center}
\end{figure}

2D point groups \cite{DH:PGSG}
%(Hestenes, 2002) 
are generated by multiplying vectors 
selected \cite{HP:crystGA,HP:sym3vec}
%(Hitzer \& Perwass 2004,2005) 
as in 
Fig. \ref{fg:2Dpg}. The index $p$ can be used to denote these groups as in Table 
\ref{tb:2Dpg}. For example the hexagonal point group is given by multiplying its
two generating vectors $\gvec{a},\gvec{b}$
\begin{align}
 6 = \{
 &\vect{a},\vect{b}, R=\vect{a}\vect{b}, R^2, R^3, R^4, R^5, R^6 \doteq -1, 
 \nonumber \\
 &\vect{a}R^2, \vect{b}R^2, \vect{a}R^4, \vect{b}R^4
 \}.
\end{align}
The rotation subgroups are denoted with bars, e.g. $\bar{6}$. 
The identities $\gvec{a}^2\doteq\gvec{b}^2\doteq1$ and $R^6\doteq-1$ directly correspond to relations
in the group presentation \cite{BS:SG}
%(Souvignier, 2007) 
of $6$. 

%%%%%%%%%%%%%%%%%%%%%%%%%%%%%%%%%%%%%%%%%%%%%%%%%%%%%%%%%%%%%%%%%%%%%%%%%
% Table of 2D Point Group Notation
%%%%%%%%%%%%%%%%%%%%%%%%%%%%%%%%%%%%%%%%%%%%%%%%%%%%%%%%%%%%%%%%%%%%%%%%%
%
\begin{table}
\tabcolsep 5pt
%\small
%\begin{center}
\caption{Geometric and international notation for 2D point groups.\label{tb:2Dpg}}
\begin{tabular}{lllllllllll}
%\hline\noalign{\smallskip}
%\topline
Crystal system&\multicolumn{2}{c}{Oblique}
&\multicolumn{2}{c}{Rectangular}
&\multicolumn{2}{c}{Trigonal}
&\multicolumn{2}{c}{Square}
&\multicolumn{2}{c}{Hexagonal}
\\
\hline
%\midline
\rule{0mm}{4mm}%
%\noalign{\smallskip}\svhline\noalign{\smallskip}
geometric & $\bar{1}$& $\bar{2}$ & 1& 2 & 3& $\bar{3}$ & 4&$\bar{4}$ & 6& $\bar{6}$
\\
international  &1& 2 &m& 2mm& 3m& 3& 4mm& 4& 6mm& 6\\
%\bottomline
\end{tabular}
%\end{center}
\end{table}

\subsection{Three dimensional point groups}

\begin{figure} 
 %\begin{center}  
   \resizebox{0.95\textwidth}{!}{\includegraphics{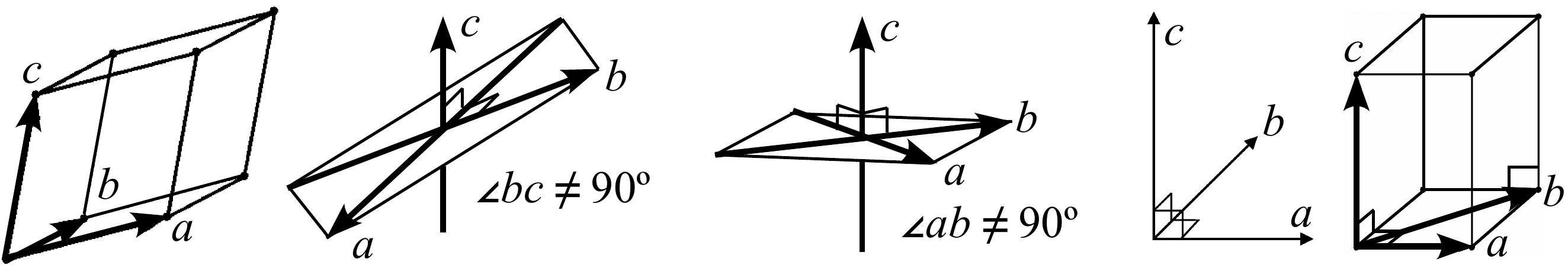}}
 %\end{center}
  \caption{From left to right: Triclinic, monoclinic inclined, 
  monoclinic orthogonal, orthorhombic, and tetragonal cell 
  vectors $\gveca,\gvecb,\gvecc$ for rod and layer groups.\label{fg:rodcells}}
\end{figure}

\begin{figure}%[ht]
 %\begin{center}
  \resizebox{0.75\textwidth}{!}{\includegraphics{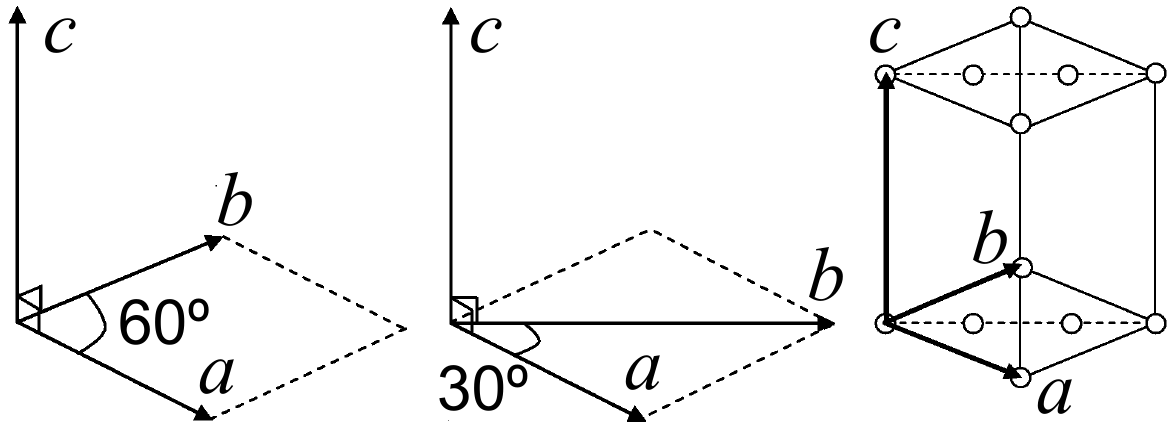}}
 %\end{center}
\caption{Generating vectors $\gveca,\gvecb,\gvecc$ of cell choices for trigonal and hexagonal crystal systems for 3D rod and layer groups. 
Left: a hexagonal primitive (\textit{hP}) cell.
%trigonal (left), 
Center: an ortho-hexagonal \textit{C}-centered cell (\textit{oC}).
Right: a hexagonal \textit{H}-centered cell (\textit{hH}) with Bravais symbol in the nomenclature: $H$ or $h$.
%hexagonally centered 
%(right, Bravais symbol: $H$ or $h$)
\label{fg:trig-hex}} 
\end{figure}

The selection of three vectors $\vect{a},\vect{b},\vect{c}$ 
from each crystal cell~\cite{DH:PGSG,HP:TheSGV}
%(Hestenes, 2002; Hitzer \& Perwass, 2006) 
for generating (cf. Table \ref{tb:3Dpg}) 3D
point groups are indicated in Figs. \ref{fg:rodcells} and \ref{fg:trig-hex}. 

%%%%%%%%%%%%%%%%%%%%%%%%%%%%%%%%%%%%%%%%%%%%%%%%%%%%%%%%%%%%%%%%%%%%%%%%%
% Table of 3D Geometric Point Group Notation
%%%%%%%%%%%%%%%%%%%%%%%%%%%%%%%%%%%%%%%%%%%%%%%%%%%%%%%%%%%%%%%%%%%%%%%%%
%
\begin{table}%[t]
%\begin{center}
\caption{\label{tb:3Dpg}%
Geometric 3D point group symbols~\cite{DH:PGSG}
%(Hestenes, 2002) 
and generators with
${\theta_{\gveca,\gvecb}=\pi/p}$,
${\theta_{\gvecb,\gvecc}=\pi/q}$,
${\theta_{\gveca,\gvecc}=\pi/2}$, $p,q\in \{1,2,3,4,6\}$, see the right side of Fig. \ref{fg:refrot}.}
\begin{tabular}{lcccccccc}
%\topline
Symbols& $p=1$ & $p\neq 1$ & $\bar{p}$ & $pq$ 
& $\bar{p}q$ & $p\bar{q}$ & $\bar{p}\bar{q}$ & $\overline{pq}$
\\
\midline
Generators &
$\gvec{a} $
& $\gvec{a},  \gvec{b}$
& $\gvec{a}\gvec{b} $ 
& $\gvec{a},  \gvec{b},  \gvec{c} $
& $\gvec{a}\gvec{b},  \gvec{c} $ 
& $\gvec{a},  \gvec{b}\gvec{c} $
& $\gvec{a}\gvec{b},  \gvec{b}\gvec{c} $ 
& $\gvec{a}\gvec{b}\gvec{c}$
\\
%\bottomline
\end{tabular}
%\end{center}
\end{table}

The selection of three vectors $\gvec{a},\gvec{b},\gvec{c}$, 
see the right side of Fig. \ref{fg:refrot},
from each crystal cell \cite{DH:PGSG,HP:crystGA,HP:sym3vec}
%(Hestenes, 2002; Hitzer \& Perwass, 2004,2005) 
for generating all 3D
point groups is indicated in Figs. \ref{fg:rodcells} and \ref{fg:trig-hex}. 
%Fig. \ref{fg:3Dcells}. 
Using $\angle(\gvec{a},\gvec{b})$ and $\angle(\gvec{b},\gvec{c})$ we can denote all 32 3D point groups (and their associated crystal classes) as in Table \ref{tb:3Dpg}. For example the monoclinic 
point groups are then 
(int. symbols of Hermann-Maugin: $2/\text{m}$, $\text{m}$ and $2$, respectively)
\begin{align}
  2\bar{2} &= \{\gvec{c}, R=\gvec{a}\wedge\gvec{b}=i\gvec{c}, i=\gvec{c}R, 1 \}, 
  \\
  1 &= \{\gvec{c}, 1 \},
  \\
  \bar{2} &= \{i\gvec{c}, 1 \}.
\end{align}

\subsection{Space groups \label{sc:SG}}

The smooth composition with translations is best done in the conformal
model \cite{SL:Diss,LL:ConfTrCA,HLR:NewAlgTools,EH:ConfM,HL:invalg}
%(Lie, 1872; Lounesto \& Latvamaa, 1980; Hestenes \textit{et al.}, 2001, Hitzer, 2004; Li, 2008)
of Euclidean space (in the GA of $\mathbb{R}^{4,1}$), 
which adds two null-vector dimensions for 
the origin $\gvec{e}_0$ and infinity $\gvec{e}_{\infty}$ such that
\begin{align}
  &X = \gvec{x} + \frac{1}{2}\gvec{x}^2\gvec{e}_{\infty}+\gvec{e}_0, 
  %\nonumber 
  \\
  &\gvec{e}_0^2 = \gvec{e}_{\infty}^2=X^2=0,
  \\
  &X\cdot \gvec{e}_{\infty} = -1.
  %\nonumber
\end{align}
The $+\gvec{e}_0$ term integrates projective geometry,
and the $+ \frac{1}{2}\gvec{x}^2\gvec{e}_{\infty}$ term
ensures $X^2=0$.
The inner product of two conformal points gives their Euclidean distance and 
therefore a plane $m$ equidistant from two points $A,B$ as
\begin{align}
  X\cdot A = -\frac{1}{2}(\gvec{x}-\gvec{a})^2 \,\,
  &\Rightarrow \,\,X\cdot (A-B)=0,
  \\
  m=A-B &\propto \gvec{p}-d\,\gvec{e}_{\infty},
\end{align}
where $\gvec{p}$ is a unit normal to the plane and $d$ its signed scalar distance 
from the origin. Reflecting at two parallel planes $m,m^{\prime}$ with 
distance $\gvec{t}/2$ we get the so-called \textit{translator} 
(\textit{transla}tion opera\textit{tor} by $\gvec{t}\,$)
\begin{equation}
  X^{\prime} = m^{\prime}m\,X\,mm^{\prime} = T_{\gvec{t}}^{-1} X T_{\gvec{t}},
  \quad T_{\gvec{t}}=1+\frac{1}{2}\gvec{t}\gvec{e}_{\infty}.
\end{equation}
Reflection at two non-parallel planes $m,m^{\prime}$ yields the rotation around
the $m,m^{\prime}$-intersection line axis by twice the angle subtended by $m,m^{\prime}$.

Group theoretically the conformal group $C(3)$ is isomorphic to $O(4,1)$ and the
Euclidean group $E(3)$ is the subgroup of $O(4,1)$ leaving infinity 
$\gvec{e}_{\infty}$ invariant \cite{DH:PGSG,HH:SGinGA,HL:invalg}.
%(Hestenes, 2002; Hestenes\& Holt, 2007; Li, 2008). 
Now general translations and rotations are represented by geometric products of vectors. 
To study combinations of versors it is useful to know that (cf. Table \ref{tb:reftr})
\begin{equation}
  T_{\gvec{t}}\,\gvec{a} = \gvec{a} \,T_{\gvec{t}^{\,\,\prime}}\,, 
  \quad \quad
  \gvec{t}^{\,\prime}=-\gvec{a}^{\,\,-1}\gvec{t}\,\gvec{a}\,.
\end{equation}
Applying these techniques one can compactly tabulate geometric space 
group symbols and generators \cite{HH:SGinGA}.
%(Hestenes \& Holt, 2007). 
Table \ref{tb:3Dsg} implements this for the 13 monoclinic space groups. 
All this is interactively visualized \cite{HP:TheSGV}
%(Hitzer \& Perwass, 2006) 
by the Space Group Visualizer \cite{HP:SGV}.
%(Perwass \& Hitzer, 2005a).

%%%%%%%%%%%%%%%%%%%%%%%%%%%%%%%%%%%%%%%%%%%%%%%%%%%%%%%%%%%%%%%%%%%%%%%%%
% Table of Computing with Refl. and Transl.
%%%%%%%%%%%%%%%%%%%%%%%%%%%%%%%%%%%%%%%%%%%%%%%%%%%%%%%%%%%%%%%%%%%%%%%%%
%
\begin{table}
%\begin{center}
\caption{\label{tb:reftr}%
Computing with reflections and translations. The vectors
$\gvec{a},\gvec{b}$ are pictured in Fig. \ref{fg:2Dpg}.}
\begin{tabular}{lccccccc}
%\topline
\rule{0mm}{3mm}%
$\angle(\gvec{a},\gvec{b})$ & $180^{\circ}$ & $90^{\circ}$ & $60^{\circ}$ 
& $45^{\circ}$ & $30^{\circ}$ 
\\
\hline
%\midline
\rule{0mm}{4mm}%
$T_{\gvec{a}}\,\gvec{b}=$ 
& $\,\,\, \gvec{b}\,T_{-\gvec{a}} \,\,\, $
& $\,\,\, \gvec{b}\,T_{\gvec{a}}  \,\,\, $ 
& $\,\,\, \gvec{b}\,T_{\gvec{a}-\gvec{b}} \,\,\, $
& $\,\,\, \gvec{b}\,T_{\gvec{a}-\gvec{b}} \,\,\, $ 
& $\,\,\, \gvec{b}\,T_{\gvec{a}-\gvec{b}} \,\,\, $
\\
\rule[-2mm]{0mm}{5.9mm} 
$T_{\gvec{b}}\,\gvec{a}=$ 
& $\,\,\, \gvec{a}\,T_{-\gvec{b}} \,\,\, $
& $\,\,\, \gvec{a}\,T_{\gvec{b}}  \,\,\, $ 
& $\,\,\, \gvec{a}\,T_{\gvec{b}-\gvec{a}}  \,\,\, $
& $\,\,\, \gvec{a}\,T_{\gvec{b}-2\gvec{a}} \,\,\, $ 
& $\,\,\, \gvec{a}\,T_{\gvec{b}-3\gvec{a}} \,\,\, $
\\
%\bottomline
\end{tabular}
%\end{center}
\end{table}

%%%%%%%%%%%%%%%%%%%%%%%%%%%%%%%%%%%%%%%%%%%%%%%%%%%%%%%%%%%%%%%%%%%%%%%%%
% Table of 3D Monoclinic SG Versors
%%%%%%%%%%%%%%%%%%%%%%%%%%%%%%%%%%%%%%%%%%%%%%%%%%%%%%%%%%%%%%%%%%%%%%%%%
%
\begin{table}
%\begin{center}
\caption{Monoclinic space group versor generators, 
$T^A=T^{1/2}_{\gvec{b}+\gvec{c}}$,  
col. 1: international space group number~\cite{TH:ITC2005},
%(Hahn, 2005), 
col. 2: international space group symbol~\cite{TH:ITC2005},
col. 3: geometric space group symbol~\cite{HH:SGinGA},
%(Hestenes \& Holt, 2007).
col. 4: geometric space group versor generators~\cite{HH:SGinGA},
cols. 5 and 6: alternative geometric space group versor generators \cite{HP:3VGenSG}.
The pure translators $T_{\gvec{a}}, T_{\gvec{b}}, T_{\gvec{c}}$ are omitted.
\label{tb:3Dsg}}
\begin{tabular}{clllll}
%\topline
%Int.\#
%& Int. name
%& Geo. name
%%~\cite{DH:PGSG} 
%& Geo. generators 
%& Int. generators &
%Alt. generators\\
\rule{0mm}{3mm}%
1. &  2. & 3. & 4. & 5. & 6. \\
\hline
%\midline
\rule{0mm}{4mm}%
%\rule{0mm}{3.7mm}%
3 & $P2$ &$P\bar{2}$ &
$i\gvec{c}=\gvec{a} \wedge \gvec{b} $ &
\\
4 & $P2_1$ & $P\bar{2}_1$ &
$i\gvec{c}T^{1/2}_{\gvec{c}}$ &
\\
5 & $C2$ &$A\bar{2}$ &
$i\gvec{c}, \,\,T^A$ &
\\
6 & $Pm$&$P1$ &
$\gvec{c}$ &
\\
7 &$Pc$& $P_a 1$ &
$\gvec{c} T^{1/2}_{\gvec{a}}$ &
\\
8 &$Cm$& $A1$ &
$\gvec{c}, \,\,T^A$ &
\\
9 &$Cc$& $A_a 1$ & 
$\gvec{c}T^{1/2}_{\gvec{a}},\,\,T^A$ &
\\
10 &$P2/m$& $P2\bar{2}$ &
$\gvec{c}, \,\,i\gvec{c}$ & 
$i, \,\,i\gvec{c}$ &
$i, \,\,\gvec{c}$
\\
11 &$P2_1/m$& $P2\bar{2}_1$ &
$\gvec{c}, \,\,i\gvec{c}T^{1/2}_{\gvec{c}}$ &
$i, \,\,i\gvec{c}T^{1/2}_{\gvec{c}}$ &
$i, \,\,\gvec{c}T^{1/2}_{\gvec{c}}$ 
\\
12 &$C2/m$& $A2\bar{2}$ &
$\gvec{c}, \,\, i\gvec{c}, \,\, T^A$ &
$i T^A, \,\,i\gvec{c}T^A, T^A$ &
$i, \,\, \gvec{c}, \,\, T^A$
\\
13 &$P2/c$& $P_a2\bar{2}$ &
$\gvec{c}T^{1/2}_{\gvec{a}}, \,\, i\gvec{c}$ &
$i, \,\,i\gvec{c}T^{1/2}_{\gvec{a}}$ &
$i, \,\,\gvec{c}T^{1/2}_{\gvec{a}}$
\\
14 &$P2_1/c$& $P_a2\bar{2}_1$ &
$\gvec{c}T^{1/2}_{\gvec{a}}, \,\,i\gvec{c} T^{1/2}_{\gvec{c}}$ &
$i, \,\,i\gvec{c}T^{1/2}_{\gvec{a}+\gvec{c}}$ &
$i, \,\, \gvec{c}T^{1/2}_{\gvec{a}+\gvec{c}}$
\\
\rule[-2mm]{0mm}{5.9mm} 
15 &$C2/c$& $A_a2\bar{2}$ &
$\gvec{c}T^{1/2}_{\gvec{a}}, \,\,i\gvec{c}, \,\, T^A$ &
$i, \,\,i\gvec{c}T^{1/2}_{\gvec{a}}, T^A$ &
$i, \,\,\gvec{c}T^{1/2}_{\gvec{a}}, T^A$
\\ 
%\bottomline 
\end{tabular}
%\end{center}
\end{table}

\section{Subperiodic groups represented in Clifford's geometric algebra 
         \label{sc:subp}}
         
Now we begin to explain the details of the new geometric algebra based representation
of so-called \textit{subperiodic} space groups. 
These include the seven \textit{frieze} groups (in 2D space, 1 degree of freedom (DOF) for translation),
the 75 \textit{rod} groups (in 3D space, 1 DOF for translation),
and the 80 \textit{layer} groups (in 3D space, 2 DOF for translations).

Compared to the geometric 2D and 3D space group symbols in \cite{HH:SGinGA}
%(Hestenes \& Holt, 2007) 
we have introduced dots: If one or two dots
occur between the Bravais symbol (\scriptp, $p$, $c$) and index $1$,
the vector $\gvec{b}$ or $\gvec{c}$, respectively, is present in the generator list. 
If one or two dots appear between the Bravais symbol and the index $2$
(without or with bar), then the vectors $\gvec{b},\gvec{c}$ or $\gvec{a},\gvec{c}$, respectively,
are present in the generator list.

In agreement~\cite{HH:SGinGA}
%(Hestenes \& Holt, 2007) 
the indexes $a,b,c,n$ (and $g$ for frieze groups) in first, 
second or third position 
after the Bravais symbol indicate that the reflections $\gvec{a},\gvec{b},\gvec{c}$ (in this order) become glide reflections. 
An index $n$ indicates diagonal glides. 
The dots also serve as symbolic $a,b,c$ position indicators. 
For example rod group $5$: \scriptp${_c1}$ has glide reflection
$\gvec{a}T_{\gvec{c}}^{1/2}$, rod group $19$: \scriptp${._c2}$ has $\gvec{b}T_{\gvec{c}}^{1/2}$, 
and layer group $39$: $p_b2_a2_n$ has $\gvec{a}{T_{\gvec{b}}}^{\frac{1}{2}}$, $\gvec{b}{T^{1/2}_{\gvec{a}}}$ 
and $\gvec{c}{T_{\gvec{a}+\gvec{b}}}^{1/2}$.

The notation $\overline{n}_p$ indicates a right handed screw rotation
of $2\pi/n$ around the $\overline{n}$-axis, with pitch $T^{p/n}_{\gvec{t}}$
where $\gvec{t}$ is the shortest lattice translation vector parallel to the axis, in
the screw direction. For example the layer group $21$: $p\bar{2}\bar{2}_1\bar{2}_1$
has the screw generators $\gvec{b}\gvec{c}T_{\gvec{a}}^{1/2}$ and $\gvec{a}\gvec{c}T_{\gvec{b}}^{1/2}$.

In the following we discuss specific issues for frieze groups,
rod groups and layer groups.

\subsection{Frieze groups}

\begin{figure}
 %\begin{center}
\resizebox{0.75\textwidth}{!}{\includegraphics{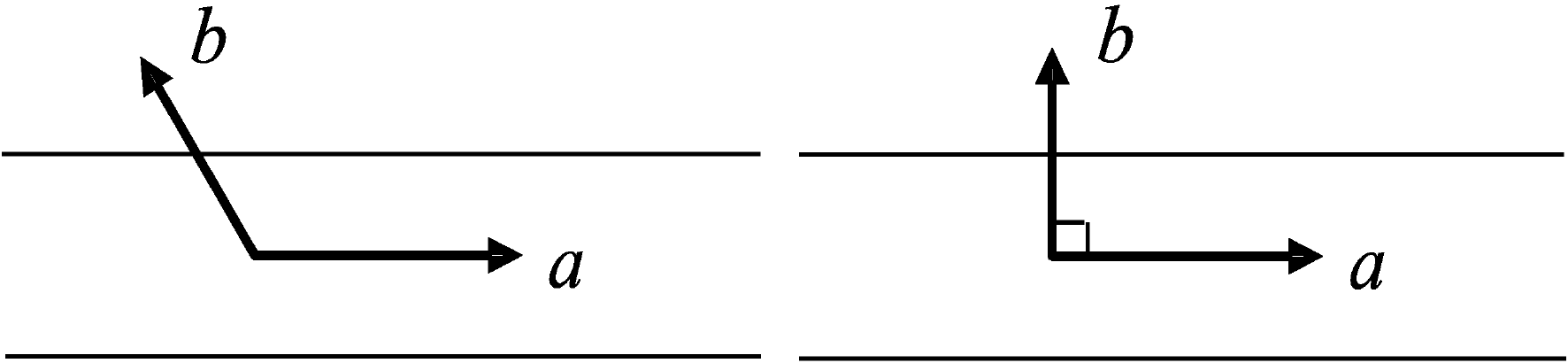}}
%resizebox{0.75\textwidth}{!}{\includegraphics*[0pt,45pt][511pt,163pt]{friezecells}}
 %\end{center}
  \caption{Generating vectors $\gveca,\gvecb$ of oblique and rectangular cells for 2D frieze groups.\label{fg:friezecells}}
\end{figure}

Figure \ref{fg:friezecells} shows the generating vectors $\gvec{a}, \gvec{b}$ of oblique and rectangular cells for 2D frieze groups. 
The only translation direction is $\gvec{a}$, frieze groups are thus subgroups of plane space groups, which can also be fully visualized with the interactive Space Group Visualizer software \cite{EH:VisPlaneG}. Table \ref{tb:frieze} lists
the seven frieze groups with \textit{new geometric symbols} and \textit{generators}.  
%The abbreviations SG\# and SGN mean space group number and space group name (symbol), respectively. 

\subsection{Rod groups}

Figure \ref{fg:rodcells} shows the generating vectors $\gvec{a}, \gvec{b}, \gvec{c}$ of 
triclinic, monoclinic, orthorhombic and tetragonal cells 
for 3D rod and layer groups. 
Figure \ref{fg:trig-hex} shows the same for trigonal and hexagonal 
cells. 
For rod groups the only translation direction is $\gvec{c}$. 
There is a total of 75 rod groups in all 3D crystal systems. 
Table \ref{tb:rodgr1} lists
the triclinic, monoclinic and orthorhombic rod groups with \textit{new geometric symbols} 
and \textit{generators}:
Rod group number (col. 1),
international rod group notation \cite{KL:ITE}
%(Kopsky \&  Litvin, 2002) 
(col. 2), 
international 3D space super group numbers \cite{TH:ITC2005}
%(Hahn, 2005) 
(col. 3),
and notation \cite{TH:ITC2005}
%(Hahn, 2005) 
(col. 4),
geometric 3D space super group notation \cite{HH:SGinGA}
%(Hestenes \& Holt, 2007) 
(col. 5),
\textit{geometric rod group notation} (col. 6),
\textit{geometric algebra generators} (col. 7).
The tetragonal and trigonal rod groups are listed in Table \ref{tb:rodg-tetra-trig}, and the hexagonal rod groups in Table \ref{tb:rodg-hex}.

Note that in the last two rows of Table \ref{tb:rodg-tetra-trig} we give in col. 5 the geometric 3D space super group notation exactly as found in \cite{HH:SGinGA}. But for full consistency with the choice of vectors in Table \ref{tb:3Dpg} and Fig. \ref{fg:trig-hex}, we have decided to modify the rod group notation (and their versor generators) specified in col. 6 (and col. 7) of the last two rows of Table \ref{tb:rodg-tetra-trig}.

\subsection{Layer groups}

For layer groups the two translation directions are $\gvec{a},\gvec{b}$. 
There is a total of 80 layer groups. 
Table \ref{tb:layer-tric-mon} lists
the triclinic and monoclinic 3D layer groups with new geometric symbols and generators: 
Layer group number (col. 1),
international layer group notation\footnote{Further well known and used layer group symbols are due to Wood, and to Bohm and Dornberger-Schiff \cite{KL:NomSubpGs}.} \cite{KL:ITE}
%(Kopsky \&  Litvin, 2002) 
(col. 2), 
international 3D space super group numbers \cite{TH:ITC2005}
%(Hahn, 2005) 
(col. 3),
and notation \cite{TH:ITC2005}
%(Hahn, 2005) 
(col. 4),
geometric 3D space super group notation \cite{HH:SGinGA}
%(Hestenes \& Holt, 2007) 
(col. 5),
\textit{geometric layer group notation} (col. 6),
\textit{geometric algebra generators} (col. 7).
Table \ref{tb:layer-orthorh} lists the orthorhombic/rectangular layer groups, and Table \ref{tb:layer-tetra-tri-hex} the tetragonal/square, trigonal/hexagonal and hexagonal/hexagonal layer groups. 
The layer groups are classified according to their 3D crystal system/2D Bravais system\footnote{Note that \textit{Bravais systems} have officially been renamed \textit{lattice systems} since 2002.}.
The monoclinic/oblique (rectangular) system corresponds to the
monoclinic/orthogonal (inclined) system of Fig. \ref{fg:rodcells}. 
Figure \ref{fg:trig-hex} shows the hexagonally centered cell 
with Bravais symbols $H$ (space group) and $h$ (layer group).

Note that we use in Table \ref{tb:layer-tetra-tri-hex} for the symmorphic space group No. 81 the rotary reflection generator $\gvec{a}\gvec{b}\gvec{c}$ and not $\gvec{b}\gvec{a}\gvec{c}$ of Table 5 of \cite{HH:SGinGA}. But the point groups and symmorphic space groups generated by $\gvec{a}\gvec{b}\gvec{c}$ and $\gvec{b}\gvec{a}\gvec{c}$ are the same, because for $p=4$ we have the following equalities (up to non-zero scalar factors, which cancel out in \eqref{eq:symtrafo})
\be
  \gvec{b}\gvec{a} \dot{=}(\gvec{a}\gvec{b})^3,
  \quad
  (\gvec{b}\gvec{a})^2\dot{=}(\gvec{a}\gvec{b})^2, 
  \quad
  (\gvec{b}\gvec{a})^3\dot{=}\gvec{a}\gvec{b},
\ee
and hence with $q=2$, $\gvec{b}\gvec{a}\gvec{c}=\gvec{c}\gvec{b}\gvec{a}$, $\gvec{a}\gvec{b}\gvec{c}=\gvec{c}\gvec{a}\gvec{b}$, and $\gvec{c}^2\dot{=}1$ we also have
\be
  \gvec{b}\gvec{a}\gvec{c}\dot{=}(\gvec{a}\gvec{b}\gvec{c})^3,
  \quad
  (\gvec{b}\gvec{a}\gvec{c})^2\dot{=}(\gvec{a}\gvec{b}\gvec{c})^2, 
  \quad
  (\gvec{b}\gvec{a}\gvec{c})^3\dot{=}\gvec{a}\gvec{b}\gvec{c}.
\ee
That is, the two sets of point transformations generated by the integer powers of the generators $\gvec{a}\gvec{b}\gvec{c}$ and $\gvec{b}\gvec{a}\gvec{c}$ are identical for $p=4$, $q=2$. We have therefore decided for consistency with Table \ref{tb:3Dpg}, to use for the space group No. 81 the geometric symbol $P\overline{42}$ and the generator $\gvec{a}\gvec{b}\gvec{c}$. A similar argument is valid for our use of the generator $\gvec{a}\gvec{b}\gvec{c}$ for space group No. 147 in our Table \ref{tb:layer-tetra-tri-hex}, instead of $\gvec{b}\gvec{a}\gvec{c}$ in Table 5 of \cite{HH:SGinGA}.

\section{Conclusion}

\begin{figure}%[hbtp]
\begin{center}
  \resizebox{0.7\textwidth}{!}{\includegraphics{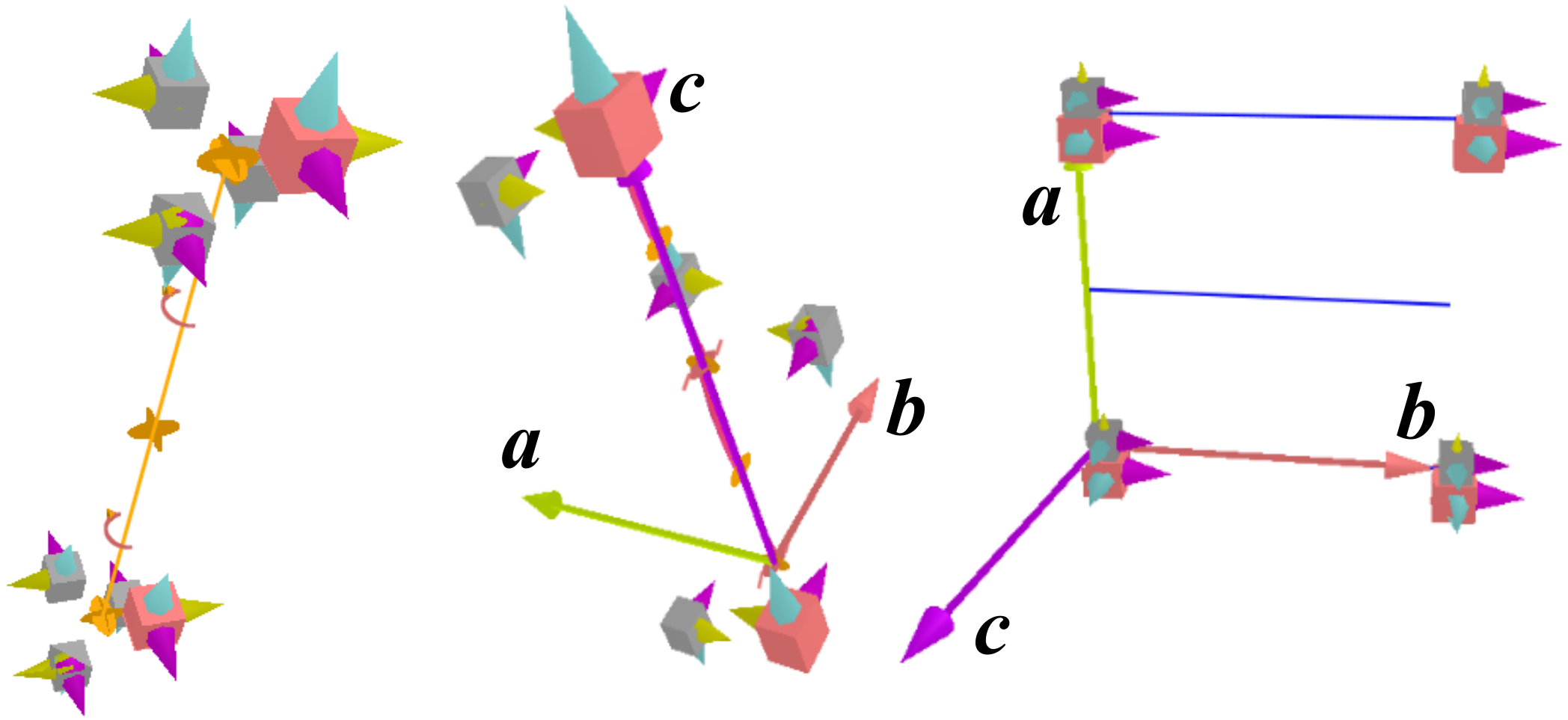}}
%\resizebox{0.8\textwidth}{!}
%{\includegraphics{twodab.eps}}
\caption{How a future subperiodic space group viewer software might depict
rod groups 13: \scriptp$\bar{2}\bar{2}\bar{2}$ 
and 14: \scriptp$\bar{2}_{1}\bar{2}\bar{2}$, 
and the layer group 11: $p1$, based on~cite{HP:TheSGV}.\label{fg:extra}}
%(Hitzer \& Perwass, 2006).}
\end{center}
\end{figure}

We have devised a new representation
for the 162 subperiodic space groups in Clifford's geometric algebra using versors (Clifford monomials, Lipschitz elements). 
In the future this may be extended to magnetic subperiodic 
space groups. 
We expect that the present work forms a suitable
foundation for interactive visualization software of subperiodic
space groups~\cite{HP:TheSGV}.
%(Hitzer \& Perwass, 2006). 
Fig. \ref{fg:extra} shows how the 
rod groups 13: \scriptp$\bar{2}\bar{2}\bar{2}$ 
and 14: \scriptp$\bar{2}_{1}\bar{2}\bar{2}$, 
and the layer group 11: $p1$ might be visualized in the future, 
based on~\cite{HP:TheSGV}.
%(Hitzer \& Perwass, 2006).

\subsection*{Acknowledgment}

E. Hitzer wishes to thank God for his wonderful creation: 
\textit{You answer us with awesome deeds of righteousness,
       O God our Savior,
       the hope of all the ends of the earth
       and of the farthest seas ...}~\cite{Psalm65:5}.
He wishes to thank 
his family for their loving support, and 
D. Hestenes, C. Perwass, M. Aroyo, D. Litvin, A. Hayashi, N. Onoda, Y. Koga, H. Zimmermann and the anonymous reviewers. We particularly thank one of the reviewers for pointing out references \cite{RAV:CACoinPL,RAAG:CLHypPlane,RAAV:AlgCartDeud,FU:GenOrthMatr,YMZ:StrCSL,YMZ:IndCIHypLat}.

%
%
%
%%%%%%%%%%%%%%%%%%%%%%%%%%%%%%%%%%%%%%%%%%%%%%%%%%%%%%%%%%%%%%%%%%%%%%%%%
% Table of frieze groups
%%%%%%%%%%%%%%%%%%%%%%%%%%%%%%%%%%%%%%%%%%%%%%%%%%%%%%%%%%%%%%%%%%%%%%%%%
%
\begin{table}
\tabcolsep 5pt
%\small
%\begin{center}
%
\caption{Table of frieze groups. 
Group number (col. 1),
international frieze group notation \cite{KL:ITE}
%(Kopsky \&  Litvin, 2002) 
(col. 2), 
international 3D space super group numbers \cite{TH:ITC2005}
%(Hahn, 2005) 
(col. 3),
and notation \cite{TH:ITC2005}
%(Hahn, 2005) 
(col. 4),
geometric 3D space super group notation \cite{HH:SGinGA}
%(Hestenes \& Holt, 2007) 
(col. 5),
international 2D space super group numbers \cite{TH:ITC2005}
%(Hahn, 2005) 
(col. 6),
and notation \cite{TH:ITC2005}
%(Hahn, 2005) 
(col. 7),
geometric 2D space super group notation \cite{HH:SGinGA}
%(Hestenes \& Holt, 2007) 
(col. 8),
\textit{geometric frieze group notation} (col. 9),
\textit{geometric algebra frieze group versor generators} (col. 10).
The pure translator $T_\gveca$ is omitted.}
\label{tb:frieze}
\begin{tabular}{cccccccccc}
%\topline
 %Frieze   & Intern. &  3D   & Intern. & Geom.  & 2D   & Intern. & Geom.  & Geom.  & Frieze Group \\
 %Group \# & Notat.  &  SG\# & 3D SGN  & 3D SGN & SG\# & 2D SGN  & 2D SGN & Notat. & Generators \\  
 \rule{0mm}{3mm}%
    1. &  2. & 3. & 4. & 5. & 6. & 7. & 8. & 9. & 10. \\
\midline
\multicolumn{10}{l}{Oblique} \\ 
\hline
%\midline
\rule{0mm}{4mm}%
${1}$ & {\scriptp}$1$   & 1 & $P1$ & $P\overline{1}$ & 1 & $p1$ & $p\overline{1}$ & \scriptp$\overline{1}$ &  \\ 
${2}$ & \scriptp$211$ & 3 & $P2$ & $P\overline{2}$ & 2 & $p2$ & $p\overline{2}$ & \scriptp$\overline{2}$ & $\gveca \wedge \gvecb$ \\
%\bottomline
\hline
\multicolumn{10}{l}{Rectangular} \\ 
\midline
${3}$ & \scriptp$1m1$ & 6  & $Pm$   & $P1$     & 3 & $pm$ & $p1$   & \scriptp1          & $\gveca$ \\
&&&&&& $(p1m1)$ &&& \\
${4}$ & \scriptp$11m$ & 6  & $Pm$   & $P1$     & 3 & $pm$ & $p1$   & \scriptp$\,.1$         & $\gvecb$ \\
&&&&&& $(p11m)$ &&& \\
${5}$ & \scriptp$11g$ & 7  & $Pc$   & $P_{a}1$ & 4 & $pg$ & $p_g1  $ & \scriptp${\,.\,}_{g}1$ & $\gvecb{T_\gveca}^{1/2}$ \\
&&&&&& $(p11g)$ &&& \\
${6}$ & \scriptp$2mm$ & 25 & $Pmm2$ & $P2$     & 6 & $p2mm$     & $p2$   & \scriptp2          & $\gveca, \gvecb$ \\
\rule[-2mm]{0mm}{5.9mm}%
${7}$ & \scriptp$2mg$ & 28 & $Pma2$ & $P2_{a}$ & 7 & $p2mg$     & $p2_{g}$ & \scriptp$2_{g}$    & $\gveca, \gvecb{T_{\gveca}}^{1/2}$ \\
%\bottomline
\end{tabular}
%\end{center}
\end{table}
%
%
%
%%%%%%%%%%%%%%%%%%%%%%%%%%%%%%%%%%%%%%%%%%%%%%%%%%%%%%%%%%%%%%%%%%%%%%%%%
% Table of rod groups I (1 - 22)
%%%%%%%%%%%%%%%%%%%%%%%%%%%%%%%%%%%%%%%%%%%%%%%%%%%%%%%%%%%%%%%%%%%%%%%%%
%
\begin{table}%[ht]
%\begin{center}
%
\caption{Table of triclinic, monoclinic and orthorhombic rod groups. 
The pure translator $T_\gvecc$ is omitted.}
\label{tb:rodgr1}
\begin{tabular}{ccccccc}
%\topline
 %Rod       & Intern. &  3D Space   & Intern. & Geom.  & Geom.  & Rod Group \\
 %Group \#  & Notat.  &  Group \#   & 3D SGN  & 3D SGN & Notat. & Generators \\ 
 \rule{0mm}{3mm}% 
1. &  2. & 3. & 4. & 5. & 6. & 7. \\ 
%\midline
%$\#$ & Int. N & $3$D \# & $3$D IN. & $3$D GN. & GRN & generators \\ 
\midline
\multicolumn{7}{l}{Triclinic} \\ 
\hline
%\midline
\rule{0mm}{4mm}%
${1}$ & \scriptp$1$ & $1$ & $P1$ & $P\bar{1}$ & \scriptp$\bar{1}$ &  \\ 
${2}$ & \scriptp$\bar{1}$ & $2$ & $P\bar{1}$ & $P\overline{22}$ & \scriptp$\overline{22}$ & $\gveca \wedge \gvecb \wedge \gvecc$ \\
\bottomline
\multicolumn{7}{l}{Monoclinic/inclined} \\ 
\hline
%\midline
\rule{0mm}{4mm}%
${3}$ & \scriptp$211$ & $3$ & $P112$ & $P\bar{2}$ &  \scriptp$\,.\,\bar{2}$ & $\gvecb \wedge \gvecc$ \\
${4}$ & \scriptp$m11$ & $6$ & $Pm$ & $P1$ & \scriptp$1$ & $\gveca$\\
${5}$ & \scriptp$c11$ & $7$ & $Pc$ & $P_{c}1$ & \scriptp$_{c}1$ & $\gveca{T_{\gvecc}^{1/2}}$ \\
${6}$ & \scriptp$2/m11$ & $10$ & $P2/m$ & $P2\bar{2}$ & \scriptp$2\bar{2}$ & $\gveca, \gvecb \wedge \gvecc$ \\
${7}$ & \scriptp$2/c11$ & $13$ & $P2/c$ & $P_{a}2\bar{2}$ & \scriptp$_{c}2\bar{2}$ & $\gveca{T_{\gvecc}^{1/2}}$, $\gvecb \wedge \gvecc$  \\
\bottomline
\multicolumn{7}{l}{Monoclinic/orthogonal} \\ 
\hline
%\midline
\rule{0mm}{4mm}%
${8}$  & \scriptp$112$       & $3$  & $P112$      & $P\bar{2}$      & \scriptp$\bar{2}$     & $\gveca \wedge \gvecb$ \\
${9}$  & \scriptp$112_{1}$   & $4$  & $P2_{1}$    & $P\bar{2}_{1}$  & \scriptp$\bar{2}_{1}$ & $(\gveca \wedge \gvecb){T_{\gvecc}^{1/2}}$ \\
${10}$ & \scriptp$11m$       & $6$  & $Pm$        & $P1$            & \scriptp$\,..1$         & $\gvecc$ \\
${11}$ & \scriptp$112/m$     & $10$ & $P2/m$      & $P\bar{2}2$     & \scriptp$\bar{2}2$    & $\gveca \wedge \gvecb, \gvecc$ \\
${12}$ & \scriptp$112_{1}/m$ & $11$ & $P2_{1}/m$  & $P\bar{2}_{1}2$ & \scriptp$\bar{2}_12$  & $(\gveca \wedge \gvecb){T_{\gvecc}^{1/2}}, \gvecc$ \\
\bottomline
\multicolumn{7}{l}{Orthorhombic} \\ 
\hline
%\midline
\rule{0mm}{4mm}%
${13}$ & \scriptp$222$ & $16$ & $P222$ & $P\bar{2}\bar{2}\bar{2}$ &  \scriptp$\bar{2}\bar{2}\bar{2}$ & $\gveca\gvecb, \gvecb\gvecc$ \\ 
${14}$ & \scriptp$222_{1}$ & $17$ & $P222_{1}$ & $P\bar{2}_{1}\bar{2}\bar{2}$ & \scriptp$\bar{2}_{1}\bar{2}\bar{2}$ & $\gveca\gvecb{T_{\gvecc}^{1/2}}, \gvecb\gvecc$ \\
${15}$ & \scriptp$mm2$ & $25$ & $Pmm2$ & $P2$ &  \scriptp$2$ & $\gveca, \gvecb$ \\
${16}$ & \scriptp$cc2$ & $27$ & $Pcc2$ & $P_{c}2_{c}$ &  \scriptp$_{c}2_{c}$ & $\gveca{T_{\gvecc}^{1/2}}, \gvecb{T_{\gvecc}^{1/2}}$ \\
${17}$ & \scriptp$mc2_{1}$ & $26$ & $Pmc2_{1}$ & $P2_{c}$ &  \scriptp$2_{c}$ & $\gveca, \gvecb{T_{\gvecc}^{1/2}}$ \\
${18}$ & \scriptp$2mm$ & $25$ & $Pmm2$ & $P2$ &  \scriptp$\,.2$ & $\gvecb, \gvecc$\\
${19}$ & \scriptp$2cm$ & $28$ & $Pma2$ & $P2_{a}$ &  \scriptp$\,._{c}2$ & $\gvecb{T_{\gvecc}^{1/2}}, \gvecc$ \\
${20}$ & \scriptp$mmm$ & $47$ & $Pmmm$ & $P22$ &  \scriptp$22$ & $\gveca, \gvecb, \gvecc$ \\
${21}$ & \scriptp$ccm$ & $49$ & $Pccm$ & $P_{c}2_{c}2$ &  \scriptp$_{c}2_{c}2$ & $\gveca{T_{\gvecc}^{1/2}}, \gvecb{T_{\gvecc}^{1/2}}, \gvecc$ \\
\rule[-2mm]{0mm}{5mm}% 
${22}$ & \scriptp$mcm$ & $51$ & $Pmma$ & $P22_{a}$ &  \scriptp$2_{c}2$ & $\gveca, \gvecb{T_{\gvecc}^{1/2}}, \gvecc$ \\
%\bottomline
\end{tabular}
%\end{center}
\end{table}
%
%
%
%%%%%%%%%%%%%%%%%%%%%%%%%%%%%%%%%%%%%%%%%%%%%%%%%%%%%%%%%%%%%%%%%%%%%%%%%
% Table of rod groups II (23 - 52)
%%%%%%%%%%%%%%%%%%%%%%%%%%%%%%%%%%%%%%%%%%%%%%%%%%%%%%%%%%%%%%%%%%%%%%%%%
%
\begin{table}%[ht]
%\begin{center}
%
\caption{Table of tetragonal and trigonal rod groups. 
The pure translator $T_\gvecc$ is omitted.}
\label{tb:rodg-tetra-trig}
\begin{tabular}{ccccccc}
%\topline
 %Rod       & Intern. &  3D Space   & Intern. & Geom.  & Geom.  & Rod Group \\
 %Group \#  & Notat.  &  Group \#   & 3D SGN  & 3D SGN & Notat. & Generators \\ 
 \rule{0mm}{3mm}%
1. &  2. & 3. & 4. & 5. & 6. & 7. \\ 
\midline
\multicolumn{7}{l}{Tetragonal} \\ 
\hline
%\midline
\rule{0mm}{4mm}%
${23}$ & \scriptp$4$ & $75$ & $P4$ & $P\bar{4}$ & \scriptp$\bar{4}$ & $\gveca\gvecb$ \\ 
${24}$ & \scriptp$4_{1}$ & $76$ & $P4_{1}$ & $P\bar{4}_{1}$ & \scriptp$\bar{4}_{1}$ & $\gveca\gvecb{T_{\gvecc}}^{\frac{1}{4}}$ \\
${25}$ & \scriptp$4_{2}$ & $77$ & $P4_{2}$ & $P\bar{4}_{2}$ &  \scriptp$\bar{4}_{2}$ & $\gveca\gvecb{T_{\gvecc}}^{\frac{1}{2}}$ \\
${26}$ & \scriptp$4_{3}$ & $78$ & $P4_{3}$ & $P\bar{4}_{3}$ & \scriptp$\bar{4}_{3}$ & $\gveca\gvecb{T_{\gvecc}}^{\frac{3}{4}}$ \\
${27}$ & \scriptp$\bar{4}$ & $81$ & $P\bar{4}$ & $P\overline{42}$ & \scriptp$\overline{42}$ & $\gveca\gvecb\gvecc$ \\
${28}$ & \scriptp$4/m$ & $83$ & $P4/m$ & $P\bar{4}2$ & \scriptp$\bar{4}2$ & $\gveca\gvecb$, $\gvecc$ \\
${29}$ & \scriptp$4_{2}/m$ & $84$ & $P4_{2}/m$ & $P\bar{4}_{2}2$ & \scriptp$\bar{4}_{2}2$ & $\gveca\gvecb{T_{\gvecc}}^{\frac{1}{2}}$, $\gvecc$ \\
${30}$ & \scriptp$422$ & $89$ & $P422$ & $P\bar{4}\bar{2}\bar{2}$ & \scriptp$\bar{4}\bar{2}\bar{2}$ & $\gveca\gvecb$, $\gvecb\gvecc$ \\
${31}$ & \scriptp$4_{1}22$ & $91$ & $P4_{1}22$ & $P\bar{4}_{1}\bar{2}\bar{2}$ & \scriptp$\bar{4}_{1}\bar{2}\bar{2}$ & $\gveca\gvecb{T_{\gvecc}}^{\frac{1}{4}}$, $\gvecb\gvecc$ \\
${32}$ & \scriptp$4_{2}22$ & $93$ & $P4_{2}22$ & $P\bar{4}_{2}\bar{2}\bar{2}$ & \scriptp$\bar{4}_{2}\bar{2}\bar{2}$ & $\gveca\gvecb{T_{\gvecc}}^{\frac{1}{2}}$, $\gvecb\gvecc$ \\
${33}$ & \scriptp$4_{3}22$ & $95$ & $P4_{3}22$ & $P\bar{4}_{3}\bar{2}\bar{2}$ & \scriptp$\bar{4}_{3}\bar{2}\bar{2}$ & $\gveca\gvecb{T_{\gvecc}}^{\frac{3}{4}}$, $\gvecb\gvecc$ \\
${34}$ & \scriptp$4mm$ & $99$ & $P4mm$ & $P4$ & \scriptp$4$ & $\gveca$, $\gvecb$ \\
${35}$ & \scriptp$4_{2}cm$ & $101$ & $P4_{2}cm$ & $P_{c}4$ &  \scriptp$_{c}4$ & $\gveca{T_{\gvecc}}^{\frac{1}{2}}$, $\gvecb$ \\ 
${36}$ & \scriptp$4cc$ & $103$ & $P4cc$ & $P_{c}4_{c}$ & \scriptp$_{c}4_{c}$ & $\gveca{T_{\gvecc}}^{\frac{1}{2}}$, $\gvecb{T_{\gvecc}}^{\frac{1}{2}}$ \\
${37}$ & \scriptp$\bar{4}m2$ & $115$ & $P\bar{4}m2$ & $P4\bar{2}$ &  \scriptp$4\bar{2}$ & $\gveca$, $\gvecb\gvecc$ \\
${38}$ & \scriptp$\bar{4}c2$ & $116$ & $P\bar{4}c2$ & $P_{c}4\bar{2}$ &  \scriptp$_{c}4\bar{2}$ & $\gveca{T_{\gvecc}}^{\frac{1}{2}}$, $\gvecb\gvecc$ \\
${39}$ & \scriptp$4/mmm$ & $123$ & $P4/mmm$ & $P42$ &  \scriptp$42$ & $\gveca$, $\gvecb$, $\gvecc$ \\
${40}$ & \scriptp$4/mcc$ & $124$ & $P4/mcc$ & $P_{c}4_{c}2$ &  \scriptp$_{c}4_{c}2$ & $\gveca{T_{\gvecc}}^{\frac{1}{2}}$, $\gvecb{T_{\gvecc}}^{\frac{1}{2}}$, $\gvecc$ \\
${41}$ & \scriptp$4_{2}/mmc$ & $131$ & $P4_{2}/mmc$ & $P4_{c}2$ &  \scriptp$4_{c}2$ & $\gveca$, $\gvecb{T_{\gvecc}}^{\frac{1}{2}}, \gvecc$ \\
\bottomline
\multicolumn{7}{l}{Trigonal} \\ 
\hline
%\midline
\rule{0mm}{4mm}%
${42}$ & \scriptp$3$ & $143$ & $P3$ & $P\bar{3}$ & \scriptp$\bar{3}$ & $\gveca\gvecb$ \\ 
${43}$ & \scriptp$3_{1}$ & $144$ & $P3_{1}$ & $P\bar{3}_{1}$ & \scriptp$\bar{3}_{1}$ & $\gveca\gvecb{T_{\gvecc}}^{\frac{1}{3}}$ \\
${44}$ & \scriptp$3_{2}$ & $145$ & $P3_{2}$ & $P\bar{3}_{2}$ &  \scriptp$\bar{3}_{2}$ & $\gveca\gvecb{T_{\gvecc}}^{\frac{2}{3}}$ \\
${45}$ & \scriptp$\bar{3}$ & $147$ & $P\bar{3}$ & $P\overline{62}$ & \scriptp$\overline{62}$ & $\gveca\gvecb\gvecc$ \\
${46}$ & \scriptp$312$ & $149$ & $P312$ & $P\bar{3}\bar{2}$ & \scriptp$\bar{3}\bar{2}$ & $\gveca\gvecb$, $\gvecb\gvecc$ \\
${47}$ & \scriptp$3_{1}12$ & $151$ & $P3_{1}12$ & $P\bar{3}_{1}\bar{2}$ & \scriptp$\bar{3}_{1}\bar{2}$ & $\gveca\gvecb{T_{\gvecc}}^{\frac{1}{3}}$, $\gvecb\gvecc$ \\
${48}$ & \scriptp$3_{2}12$ & $153$ & $P3_{2}12$ & $P\bar{3}_{2}\bar{2}$ & \scriptp$\bar{3}_{2}\bar{2}$ & $\gveca\gvecb{T_{\gvecc}}^{\frac{2}{3}}$, $\gvecb\gvecc$ \\
${49}$ & \scriptp$3m1$ & $156$ & $P3m1$ & $P3$ & \scriptp$3$ & $\gveca$, $\gvecb$ \\
${50}$ & \scriptp$3c1$ & $158$ & $P3c1$ & $P_{c}3_{c}$ & \scriptp$_{c}3_{c}$ & $\gveca{T_{\gvecc}}^{\frac{1}{2}}$, $\gvecb{T_{\gvecc}}^{\frac{1}{2}}$ \\
${51}$ & \scriptp$\bar{3}1m$ & $162$ & $P\bar{3}1m$ & $P\bar{2}6$     & \scriptp$6\bar{2}$       & $\gveca$, $\gvecb\gvecc$ \\
\rule[-2mm]{0mm}{5mm}%
${52}$ & \scriptp$\bar{3}1c$ & $163$ & $P\bar{3}1c$ & $P\bar{2}_{c}6$ & \scriptp${_{c}}6\bar{2}$ & $\gveca{T_{\gvecc}}^{\frac{1}{2}}$, $\gvecb\gvecc$ \\
%\bottomline
\end{tabular}
%\end{center}
\end{table}
%
%
%
%%%%%%%%%%%%%%%%%%%%%%%%%%%%%%%%%%%%%%%%%%%%%%%%%%%%%%%%%%%%%%%%%%%%%%%%%
% Table of rod groups III (53 - 75)
%%%%%%%%%%%%%%%%%%%%%%%%%%%%%%%%%%%%%%%%%%%%%%%%%%%%%%%%%%%%%%%%%%%%%%%%%
%
\begin{table}
%\begin{center}
%
\caption{Table of hexagonal rod groups. 
The pure translator $T_\gvecc$ is omitted.}
\label{tb:rodg-hex}
\begin{tabular}{ccccccc}
%\topline
 %Rod       & Intern. &  3D Space   & Intern. & Geom.  & Geom.  & Rod Group \\
 %Group \#  & Notat.  &  Group \#   & 3D SGN  & 3D SGN & Notat. & Generators \\ 
 \rule{0mm}{3mm}%
1. &  2. & 3. & 4. & 5. & 6. & 7. \\ 
\hline
%\midline
\rule{0mm}{4mm}%
${53}$ & \scriptp$6$ & $168$ & $P6$ & $P\bar{6}$ & \scriptp$\bar{6}$ & $\gveca\gvecb$ \\ 
${54}$ & \scriptp$6_{1}$ & $169$ & $P6_{1}$ & $P\bar{6}_{1}$ & \scriptp$\bar{6}_{1}$ & $\gveca\gvecb{T_{\gvecc}}^{\frac{1}{6}}$ \\
${55}$ & \scriptp$6_{2}$ & $171$ & $P6_{2}$ & $P\bar{6}_{2}$ & \scriptp$\bar{6}_{2}$ & $\gveca\gvecb{T_{\gvecc}}^{\frac{1}{3}}$ \\
${56}$ & \scriptp$6_{3}$ & $173$ & $P6_{3}$ & $P\bar{6}_{3}$ & \scriptp$\bar{6}_{3}$ & $\gveca\gvecb{T_{\gvecc}}^{\frac{1}{2}}$ \\
${57}$ & \scriptp$6_{4}$ & $172$ & $P6_{4}$ & $P\bar{6}_{4}$ & \scriptp$\bar{6}_{4}$ & $\gveca\gvecb{T_{\gvecc}}^{\frac{2}{3}}$ \\
${58}$ & \scriptp$6_{5}$ & $170$ & $P6_{5}$ & $P\bar{6}_{5}$ & \scriptp$\bar{6}_{5}$ & $\gveca\gvecb{T_{\gvecc}}^{\frac{5}{6}}$ \\
${59}$ & \scriptp$\bar{6}$ & $174$ & $P\bar{6}$ & $P\bar{3}2$ & \scriptp$\bar{3}2$ & $\gveca\gvecb$, $\gvecc$ \\
${60}$ & \scriptp$6/m$ & $175$ & $P6/m$ & $P\bar{6}2$ & \scriptp$\bar{6}2$ & $\gveca\gvecb$, $\gvecc$ \\
${61}$ & \scriptp$6_{3}/m$ & $176$ & $P6_{3}/m$ & $P\bar{6}_{3}2$ & \scriptp$\bar{6}_{3}2$ & $\gveca\gvecb{T_{\gvecc}}^{\frac{1}{2}}$, $\gvecc$ \\
${62}$ & \scriptp$622$ & $177$ & $P622$ & $P\bar{6}\bar{2}$ & \scriptp$\bar{6}\bar{2}$ & $\gveca\gvecb$, $\gvecb\gvecc$ \\
${63}$ & \scriptp$6_{1}22$ & $178$ & $P6_{1}22$ & $P\bar{6}_{1}\bar{2}$ & \scriptp$\bar{6}_{1}\bar{2}$ & $\gveca\gvecb{T_{\gvecc}}^{\frac{1}{6}}$, $\gvecb\gvecc$ \\
${64}$ & \scriptp$6_{2}22$ & $180$ & $P6_{2}22$ & $P\bar{6}_{2}\bar{2}$ & \scriptp$\bar{6}_{2}\bar{2}$ & $\gveca\gvecb{T_{\gvecc}}^{\frac{1}{3}}$, $\gvecb\gvecc$ \\ 
${65}$ & \scriptp$6_{3}22$ & $182$ & $P6_{3}22$ & $P\bar{6}_{3}\bar{2}$ & \scriptp$\bar{6}_{3}\bar{2}$ & $\gveca\gvecb{T_{\gvecc}}^{\frac{1}{2}}$, $\gvecb\gvecc$ \\
${66}$ & \scriptp$6_{4}22$ & $181$ & $P6_{4}22$ & $P\bar{6}_{4}\bar{2}$ & \scriptp$\bar{6}_{4}\bar{2}$ & $\gveca\gvecb{T_{\gvecc}}^{\frac{2}{3}}$, $\gvecb\gvecc$ \\
${67}$ & \scriptp$6_{5}22$ & $179$ & $P6_{5}22$ & $P\bar{6}_{5}\bar{2}$ & \scriptp$\bar{6}_{5}\bar{2}$ & $\gveca\gvecb{T_{\gvecc}}^{\frac{5}{6}}$, $\gvecb\gvecc$ \\
${68}$ & \scriptp$6mm$ & $183$ & $P6mm$ & $P6$ & \scriptp$6$ & $\gveca$, $\gvecb$ \\
${69}$ & \scriptp$6cc$ & $184$ & $P6cc$ & $P_{c}6_{c}$ & \scriptp$_{c}6_{c}$ & $\gveca{T_{\gvecc}}^{\frac{1}{2}}$, $\gvecb{T_{\gvecc}}^{\frac{1}{2}}$ \\
${70}$ & \scriptp$6_{3}cm$ & $185$ & $P6_{3}cm$ & $P_{c}6$ & \scriptp$_{c}6$ & $\gveca{T_{\gvecc}}^{\frac{1}{2}}$, $\gvecb$ \\
${71}$ & \scriptp$\bar{6}m2$ & $187$ & $P\bar{6}m2$ & $P32$ & \scriptp$32$ & $\gveca$, $\gvecb$, $\gvecc$ \\
${72}$ & \scriptp$\bar{6}c2$ & $188$ & $P\bar{6}c2$ & $P_{c}3_{c}2$ & \scriptp$_{c}3_{c}2$ & $\gveca{T_{\gvecc}}^{\frac{1}{2}}$, $\gvecb{T_{\gvecc}}^{\frac{1}{2}}$, $\gvecc$ \\
${73}$ & \scriptp$6/mmm$ & $191$ & $P6/mmm$ & $P62$ & \scriptp$62$ & $\gveca$, $\gvecb$, $\gvecc$ \\
${74}$ & \scriptp$6/mcc$ & $192$ & $P6/mcc$ & $P_{c}6_{c}2$ & \scriptp$_{c}6_{c}2$ & $\gveca{T_{\gvecc}}^{\frac{1}{2}}$, $\gvecb{T_{\gvecc}}^{\frac{1}{2}}$, $\gvecc$ \\
\rule[-2mm]{0mm}{5mm}%
${75}$ & \scriptp$6_{3}/mcm$ & $193$ & $P6_{3}/mcm$ & $P_{c}62$ & \scriptp$_{c}62$ & $\gveca{T_{\gvecc}}^{\frac{1}{2}}$, $\gvecb$, $\gvecc$ \\
%\bottomline
\end{tabular}
%\end{center}
\end{table}
%
%
%
%%%%%%%%%%%%%%%%%%%%%%%%%%%%%%%%%%%%%%%%%%%%%%%%%%%%%%%%%%%%%%%%%%%%%%%%%
% Table of layer groups I (1 - 18)
%%%%%%%%%%%%%%%%%%%%%%%%%%%%%%%%%%%%%%%%%%%%%%%%%%%%%%%%%%%%%%%%%%%%%%%%%
%
\begin{table}%[ht]
%\begin{center}
%
\caption{Table of triclinic and monoclinic layer groups.
The pure translators $T_\gveca, T_\gvecb$ are omitted.}
\label{tb:layer-tric-mon}
\begin{tabular}{ccccccc}
%\topline
 %Layer       & Intern. &  3D Space   & Intern. & Geom.  & Geom.  & Layer Group \\
 %Group \#    & Notat.  &  Group \#   & 3D SGN  & 3D SGN & Notat. & Generators  \\ 
 \rule{0mm}{3mm}%
1. &  2. & 3. & 4. & 5. & 6. & 7. \\ 
\midline
\multicolumn{7}{l}{Triclinic/oblique} \\ 
\hline
%\midline
\rule{0mm}{4mm}%
${1}$ & $p1$ & $1$ & $P1$ & $P\bar{1}$ & $p\bar{1}$ &  \\ 
${2}$ & $p\bar{1}$ & $2$ & $P\bar{1}$ & $P\overline{22}$ & $p\overline{22}$ & $\gveca \wedge \gvecb \wedge \gvecc$ \\ 
\bottomline
\multicolumn{7}{l}{Monoclinic/oblique} \\
\hline
%\midline
\rule{0mm}{4mm}%
${3}$ & $p112$ & $3$ & $P2$ & $P\bar{2}$ &  $p\bar{2}$ & $\gveca\wedge \gvecb$ \\ 
${4}$ & $p11m$ & $6$ & $Pm$ & $P1$ &  $p..1$ & $\gvecc$ \\
${5}$ & $p11a$ & $7$ & $Pc$ & $P_{a}1$ &  $p.._{a}1$ & $\gvecc{T_{\gveca}}^{\frac{1}{2}}$ \\
${6}$ & $p112/m$ & $10$ & $P2/m$ & $P\bar{2}2$ &  $p\bar{2}2$ & $\gveca\wedge \gvecb$, $\gvecc$ \\
${7}$ & $p112/a$ & $13$ & $P2/c$ & $P_{a}2\bar{2}$ &  $p\bar{2}2_{a}$ & $\gveca\wedge \gvecb$, $\gvecc{T_{\gveca}}^{\frac{1}{2}}$  \\
\bottomline
\multicolumn{7}{l}{Monoclinic/rectangular} \\
\hline
%\midline
\rule{0mm}{4mm}%
${8}$ & $p211$ & $3$ & $P2$ & $P\bar{2}$ &  $p.\bar{2}$ & $\gvecb\wedge \gvecc$ \\
${9}$ & $p2_{1}11$ & $4$ & $P2_{1}$ & $P\bar{2}_{1}$ &  $p.\bar{2}_{1}$ & $(\gvecb\wedge \gvecc){T_{\gveca}}^{\frac{1}{2}}$\\
${10}$ & $c211$ & $5$ & $C2$ & $A\bar{2}$ &  $c.\bar{2}$ & $\gvecb\wedge \gvecc$, ${T^{1/2}_{\gveca+\gvecb}}$\\
${11}$ & $pm11$ & $6$ & $Pm$ & $P1$ &  $p1$ & $\gveca$\\
${12}$ & $pb11$ & $7$ & $Pc$ & $P_{a}1$ &  $p_{b}1$ & $\gveca{T_{\gvecb}}^{\frac{1}{2}}$\\
${13}$ & $cm11$ & $8$ & $Cm$ & $A1$ &  $c1$ & $\gveca$, ${T^{1/2}_{\gveca+\gvecb}}$\\
${14}$ & $p2/m11$ & $10$ & $P2/m$ & $P2\bar{2}$ &  $p2\bar{2}$ & $\gveca$, $\gvecb\wedge \gvecc$\\
${15}$ & $p2_{1}/m11$ & $11$ & $P2_{1}/m$ & $P2\bar{2}_{1}$ &  $p2\bar{2}_{1}$ & $\gveca$, $(\gvecb\wedge \gvecc){T_{\gveca}}^{\frac{1}{2}}$\\
${16}$ & $p2/b11$ & $13$ & $P2/c$ & $P_{a}2\bar{2}$ &  $p_{b}2\bar{2}$ & $\gveca{T_{\gvecb}}^{\frac{1}{2}}$, $\gvecb\wedge \gvecc$\\
${17}$ & $p2_{1}/b11$ & $14$ & $P2_{1}/c$ & $P_{a}2\bar{2}_{2}$ &  $p_{b}2\bar{2}_{1}$ & $\gveca{T_{\gvecb}}^{\frac{1}{2}}$, $(\gvecb\wedge \gvecc){T_{\gveca}}^{\frac{1}{2}}$\\
\rule[-2mm]{0mm}{5mm}%
${18}$ & $c2/m11$ & $12$ & $C2/m$ & $A2\bar{2}$ &  $c2\bar{2}$ & $\gveca$, $\gvecb\wedge \gvecc$, ${T^{1/2}_{\gveca+\gvecb}}$\\
%\bottomline
\end{tabular}
%\end{center}
\end{table}
%
%
%
%%%%%%%%%%%%%%%%%%%%%%%%%%%%%%%%%%%%%%%%%%%%%%%%%%%%%%%%%%%%%%%%%%%%%%%%%
% Table of layer groups II (19 - 48)
%%%%%%%%%%%%%%%%%%%%%%%%%%%%%%%%%%%%%%%%%%%%%%%%%%%%%%%%%%%%%%%%%%%%%%%%%
%
\begin{table}
\tabcolsep 5pt
%\small
%\begin{center}
%
\caption{Table of orthorhombic/rectangular layer groups.
The pure translators $T_\gveca, T_\gvecb$ are omitted.}
\label{tb:layer-orthorh}
\begin{tabular}{ccccccc}
%\topline
 %Layer       & Intern. &  3D Space   & Intern. & Geom.  & Geom.  & Layer Group \\
 %Group \#    & Notat.  &  Group \#   & 3D SGN  & 3D SGN & Notat. & Generators  \\  
 \rule{0mm}{3mm}%
 1. &  2. & 3. & 4. & 5. & 6. & 7. \\
\hline
%\midline
\rule{0mm}{4mm}%
${19}$ & $p222$ & $16$ & $P222$ & $P\bar{2}\bar{2}\bar{2}$ & $p\bar{2}\bar{2}\bar{2}$ & $\gveca\gvecb$, $\gvecb\gvecc$ \\ 
${20}$ & $p2_{1}22$ & $17$ & $P222_{1}$ & $P\bar{2}_{1}\bar{2}\bar{2}$ & $p\bar{2}\bar{2}_{1}\bar{2}$ & $\gveca\gvecb$, $\gvecb\gvecc{T_{\gveca}}^{\frac{1}{2}}$ \\ 
${21}$ & $p2_{1}2_{1}2$ & $18$ & $P2_{1}2_{1}2$ & $P\bar{2}_{1}\bar{2}_{1}\bar{2}$ & $p\bar{2}\bar{2}_{1}\bar{2}_{1}$ & $\gvecb\gvecc{T_{\gveca}}^{\frac{1}{2}}$, $\gveca\gvecc{T_{\gvecb}}^{\frac{1}{2}}$ \\ 
${22}$ & $c222$ & $21$ & $C222$ & $C\bar{2}\bar{2}\bar{2}$ & $c\bar{2}\bar{2}\bar{2}$ & $\gveca\gvecb$, $\gvecb\gvecc$, ${T_{\gveca+\gvecb}}^{\frac{1}{2}}$ \\ 
${23}$ & $pmm2$ & $25$ & $Pmm2$ & $P2$ & $p2$ & $\gveca$, $\gvecb$ \\ 
${24}$ & $pma2$ & $28$ & $Pma2$ & $P2_{a}$ & $p2_{a}$ & $\gveca$, $\gvecb{T_{\gveca}}^{\frac{1}{2}}$ \\ 
${25}$ & $pba2$ & $32$ & $Pba2$ & $P_{b}2_{a}$ & $p_{b}2_{a}$ & $\gveca{T_{\gvecb}}^{\frac{1}{2}}$, $\gvecb{T_{\gveca}}^{\frac{1}{2}}$ \\ 
${26}$ & $cmm2$ & $35$ & $Cmm2$ & $C2$ & $c2$ & $\gveca$, $\gvecb$, ${T^{1/2}_{\gveca+\gvecb}}$ \\ 
${27}$ & $pm2m$ & $25$ & $Pmm2$ & $P2$ & $p..2$ & $\gveca$, $\gvecc$ \\ 
${28}$ & $pm2_{1}b$ & $26$ & $Pmc2_{1}$ & $P2_{c}$ & $p.._{b}2$ & $\gveca$, $\gvecc{T_{\gvecb}}^{\frac{1}{2}}$ \\ 
${29}$ & $pb2_{1}m$ & $26$ & $Pmc2_{1}$ & $P2_{c}$ & $p_{b}..2$ & $\gveca{T_{\gvecb}}^{\frac{1}{2}}$, $\gvecc$ \\ 
${30}$ & $pb2b$ & $27$ & $Pcc2$ & $P_{c}2_{c}$ & $p_{b}.._{b}2$ & $\gveca{T_{\gvecb}}^{\frac{1}{2}}$, $\gvecc{T_{\gvecb}}^{\frac{1}{2}}$ \\ 
${31}$ & $pm2a$ & $28$ & $Pma2$ & $P2_{a}$ & $p.._{a}2$ & $\gveca$, $\gvecc{T_{\gveca}}^{\frac{1}{2}}$ \\ 
${32}$ & $pm2_{1}n$ & $31$ & $Pmn2_{1}$ & $P2_{n}$ & $p.._{n}2$ & $\gveca$, $\gvecc{T^{1/2}_{\gveca+\gvecb}}$ \\ 
${33}$ & $pb2_{1}a$ & $29$ & $Pca2_{1}$ & $P_{c}2_{a}$ & $p_{b}.._{a}2$ & $\gveca{T_{\gvecb}}^{\frac{1}{2}}$, $\gvecc{T_{\gveca}}^{\frac{1}{2}}$ \\ 
${34}$ & $pb2n$ & $30$ & $Pnc2$ & $P_{n}2_{c}$ & $p_{b}.._{n}2$ & $\gveca{T_{\gvecb}}^{\frac{1}{2}}$, $\gvecc{T^{1/2}_{\gveca+\gvecb}}$ \\ 
${35}$ & $cm2m$ & $35$ & $Cmm2$ & $C2$ & $c..2$ & $\gveca$, $\gvecc$, ${T^{1/2}_{\gveca+\gvecb}}$ \\ 
${36}$ & $cm2e$ & $39$ & $Aem2$ & $A_{b}2$ & $c.._{a}2$ & $\gveca$, $\gvecc{T_{\gveca}}^{\frac{1}{2}}$, ${T^{1/2}_{\gveca+\gvecb}}$ \\ 
${37}$ & $pmmm$ & $47$ & $Pmmm$ & $P22$ & $p22$ & $\gveca$, $\gvecb$, $\gvecc$ \\ 
${38}$ & $pmaa$ & $49$ & $Pccm$ & $P_{c}2_{c}2$ & $p2_{a}2_{a}$ & $\gveca$, $\gvecb{T_{\gveca}}^{\frac{1}{2}}$, $\gvecc{T_{\gveca}}^{\frac{1}{2}}$ \\ 
${39}$ & $pban$ & $50$ & $Pban$ & $P_{b}2_{a}2_{n}$ & $p_{b}2_{a}2_{n}$ & $\gveca{T_{\gvecb}}^{\frac{1}{2}}$, $\gvecb{T^{1/2}_{\gveca}}$, $\gvecc{T_{\gveca+\gvecb}}^{1/2}$ \\ 
${40}$ & $pmam$ & $51$ & $Pmma$ & $P22_{a}$ & $p2_{a}2$ & $\gveca$, $\gvecb{T_{\gveca}}^{\frac{1}{2}}$, $\gvecc$ \\ 
${41}$ & $pmma$ & $51$ & $Pmma$ & $P22_{a}$ & $p22_{a}$ & $\gveca$, $\gvecb$, $\gvecc{T_{\gveca}}^{\frac{1}{2}}$ \\ 
${42}$ & $pman$ & $53$ & $Pmna$ & $P2_{n}2_{a}$ & $p2_{a}2_{n}$ & $\gveca$, $\gvecb{T_{\gveca}}^{\frac{1}{2}}$, $\gvecc{T^{1/2}_{\gveca+\gvecb}}$ \\ 
${43}$ & $pbaa$ & $54$ & $Pcca$ & $P_{c}2_{c}2_{a}$ & $p_{b}2_{a}2_{a}$ & $\gveca{T_{\gvecb}}^{\frac{1}{2}}$, $\gvecb{T_{\gveca}}^{\frac{1}{2}}$, $\gvecc{T_{\gveca}}^{\frac{1}{2}}$ \\ 
${44}$ & $pbam$ & $55$ & $Pbam$ & $P_{b}2_{a}2$ & $p_{b}2_{a}2$ & $\gveca{T_{\gvecb}}^{\frac{1}{2}}$, $\gvecb{T_{\gveca}}^{\frac{1}{2}}$, $\gvecc$ \\ 
${45}$ & $pbma$ & $57$ & $Pbcm$ & $P_{b}2_{c}2$ & $p_{b}22_{a}$ & $\gveca{T_{\gvecb}}^{\frac{1}{2}}$, $\gvecb$, $\gvecc{T^{1/2}_{\gveca}}$ \\ 
${46}$ & $pmmn$ & $59$ & $Pmmn$ & $P22_{n}$ & $p22_{n}$ & $\gveca$, $\gvecb$, $\gvecc{T_{\gveca+\gvecb}}^{\frac{1}{2}}$ \\ 
${47}$ & $cmmm$ & $65$ & $Cmmm$ & $C22$ & $c22$ & $\gveca$, $\gvecb$, $\gvecc$, ${T^{1/2}_{\gveca+\gvecb}}$ \\ 
\rule[-2mm]{0mm}{5mm}%
${48}$ & $cmme$ & $67$ & $Cmme$ & $C22_{a}$ & $c22_{a}$ & $\gveca$, $\gvecb$, $\gvecc{T_{\gveca}}^{\frac{1}{2}}$, ${T^{1/2}_{\gveca+\gvecb}}$ \\ 
%\bottomline
\end{tabular}
%\end{center}
\end{table}
%
%
%
%%%%%%%%%%%%%%%%%%%%%%%%%%%%%%%%%%%%%%%%%%%%%%%%%%%%%%%%%%%%%%%%%%%%%%%%%
% Table of layer groups III (49 - 80)
%%%%%%%%%%%%%%%%%%%%%%%%%%%%%%%%%%%%%%%%%%%%%%%%%%%%%%%%%%%%%%%%%%%%%%%%%
%
\begin{table}
\tabcolsep 5pt
%\small
%\begin{center}
%
\caption{Table of tetragonal, trigonal and hexagonal layer groups. 
Layer groups ${57}$, ${58}$ and ${71}$ use special vector notation.
The pure translators $T_\gveca, T_\gvecb$ are omitted.}
\label{tb:layer-tetra-tri-hex}
\begin{tabular}{ccccccc}
%\topline
 %Layer       & Intern. &  3D Space   & Intern. & Geom.  & Geom.  & Layer Group \\
 %Group \#    & Notat.  &  Group \#   & 3D SGN  & 3D SGN & Notat. & Generators  \\  
 \rule{0mm}{3mm}%
 1. &  2. & 3. & 4. & 5. & 6. & 7. \\
\midline
\multicolumn{7}{l}{Tetragonal/square} \\ 
\hline
%\midline
\rule{0mm}{4mm}%
${49}$ & $p4$ & $75$ & $P4$ & $P\bar{4}$ & $p\bar{4}$ & $\gveca\gvecb$ \\ 
${50}$ & $p\bar{4}$ & $81$ & $P\bar{4}$ & $P\overline{42}$ & $p\overline{42}$ & $\gveca\gvecb\gvecc$ \\ 
${51}$ & $p4/m$ & $83$ & $P4/m$ & $P\bar{4}2$ & $p\bar{4}2$ & $\gveca\gvecb$, $\gvecc$ \\
${52}$ & $p4/n$ & $85$ & $P4/n$ & $P\bar{4}_{n}2$ & $p\bar{4}_{n}2$ & $\gveca\gvecb$, $\gvecc{T_{\gvecb}}^{\frac{1}{2}}$ \\
${53}$ & $p422$ & $89$ & $P422$ & $P\bar{4}\bar{2}\bar{2}$ & $p\bar{4}\bar{2}\bar{2}$ & $\gveca\gvecb$, $\gvecb\gvecc$ \\
${54}$ & $p42_{1}2$ & $90$ & $P42_{1}2$ & $P\bar{4}\bar{2}_{1}\bar{2}$ & $p\bar{4}\bar{2}_{1}\bar{2}$ & $\gveca\gvecb$, $\gvecb\gvecc{T^{1/2}_{2\gveca-\gvecb}}$ \\
${55}$ & $p4mm$ & $99$ & $P4mm$ & $P4$ & $p4$ & $\gveca$, $\gvecb$ \\
${56}$ & $p4bm$ & $100$ & $P4bm$ & $P_{b}4$ & $p_{b}4$ & $\gveca{T^{1/2}_{\gveca-\gvecb}}$, $\gvecb$ \\
${57}$ & $p\bar{4}2m$ & $111$ & $P\bar{4}2m$ & $P\bar{2}4$ & $p\bar{2}4$ & $\gveca\gvecc$, $\gvecb$ \\
${58}$ & $p\bar{4}2_{1}m$ & $113$ & $P\bar{4}2_{1}m$ & $P\bar{2}_{1}4$ & $p\bar{2}_{1}4$ & $\gveca\gvecc{T^{1/2}_{\gveca-\gvecb}}$, $\gvecb$ \\
${59}$ & $p\bar{4}m2$ & $115$ & $P\bar{4}m2$ & $P4\bar{2}$ & $p4\bar{2}$ & $\gveca$, $\gvecb\gvecc$ \\
${60}$ & $p\bar{4}b2$ & $117$ & $P\bar{4}b2$ & $P_{b}4\bar{2}$ & $p_{b}4\bar{2}$ & $\gveca{T^{1/2}_{\gveca-\gvecb}}$, $\gvecb\gvecc$ \\
${61}$ & $p4/mmm$ & $123$ & $P4/mmm$ & $P42$ &  $p42$ & $\gveca$, $\gvecb$, $\gvecc$ \\
${62}$ & $p4/nbm$ & $125$ & $P4/nbm$ & $P_{b}42_{n}$ &  $p_{b}42_{n}$ & $\gveca{T^{1/2}_{\gveca-\gvecb}}$, $\gvecb$, $\gvecc{T_{\gvecb}}^{\frac{1}{2}}$ \\
${63}$ & $p4/mbm$ & $127$ & $P4/mbm$ & $P_{b}42$ &  $p_{b}42$ & $\gveca{T^{1/2}_{\gveca-\gvecb}}$, $\gvecb$, $\gvecc$ \\
${64}$ & $p4/nmm$ & $129$ & $P4/nmm$ & $P42_{n}$ &  $p42_{n}$ & $\gveca$, $\gvecb$, $\gvecc{T_{\gvecb}}^{\frac{1}{2}}$ \\
\bottomline
\multicolumn{7}{l}{Trigonal/hexagonal} \\
\hline
%\midline
\rule{0mm}{4mm}%
${65}$ & $p3$ & $143$ & $P3$ & $P\bar{3}$ & $p\bar{3}$ & $\gveca\gvecb$ \\ 
${66}$ & $p\bar{3}$ & $147$ & $P\bar{3}$ & $P\overline{62}$ & $p\overline{62}$ & $\gveca\gvecb\gvecc$ \\
${67}$ & $p312$ & $149$ & $P312$ & $P\bar{3}\bar{2}$ & $p\bar{3}\bar{2}$ & $\gveca\gvecb$, $\gvecb\gvecc$ \\
${68}$ & $p321$ & $150$ & $P321$ & $H\bar{3}\bar{2}$ & $h\bar{3}\bar{2}$ & $\gveca\gvecb$, $\gvecb\gvecc$ \\
${69}$ & $p3m1$ & $156$ & $P3m1$ & $P3$ & $p3$ & $\gveca$, $\gvecb$ \\
${70}$ & $p31m$ & $157$ & $P31m$ & $H3$ & $h3$ & $\gveca$, $\gvecb$ \\
${71}$ & $p\bar{3}1m$ & $162$ & $P\bar{3}1m$ & $P\bar{2}6$ & $p\bar{2}6$ & $\gveca\gvecc$, $\gvecb$ \\
${72}$ & $p\bar{3}m1$ & $164$ & $P\bar{3}m1$ & $P6\bar{2}$ & $p6\bar{2}$ & $\gveca$, $\gvecb\gvecc$ \\
\bottomline
\multicolumn{7}{l}{Hexagonal/hexagonal} \\
\hline
%\midline
\rule{0mm}{4mm}%
${73}$ & $p6$ & $168$ & $P6$ & $P\bar{6}$ & $p\bar{6}$ & $\gveca\gvecb$ \\ 
${74}$ & $p\bar{6}$ & $174$ & $P\bar{6}$ & $P\bar{3}2$ & $p\bar{3}2$ & $\gveca\gvecb$, $\gvecc$ \\
${75}$ & $p6/m$ & $175$ & $P6/m$ & $P\bar{6}2$ & $p\bar{6}2$ & $\gveca\gvecb$, $\gvecc$ \\
${76}$ & $p622$ & $177$ & $P622$ & $P\bar{6}\bar{2}$ & $p\bar{6}\bar{2}$ & $\gveca\gvecb$, $\gvecb\gvecc$ \\
${77}$ & $p6mm$ & $183$ & $P6mm$ & $P6$ & $p6$ & $\gveca$, $\gvecb$ \\
${78}$ & $p\bar{6}m2$ & $187$ & $P\bar{6}m2$ & $P32$ & $p32$ & $\gveca$, $\gvecb$, $\gvecc$ \\
${79}$ & $p\bar{6}2m$ & $189$ & $P\bar{6}2m$ & $H32$ & $h32$ & $\gveca$, $\gvecb$, $\gvecc$ \\
\rule[-2mm]{0mm}{5mm}%
${80}$ & $p6/mmm$ & $191$ & $P6/mmm$ & $P62$ & $p62$ & $\gveca$, $\gvecb$, $\gvecc$ \\
%\bottomline
\end{tabular}
%\end{center}
\end{table}

% ------------------------------------------------------------------------
\end{document}